\documentclass[structabstract]{aa}
\usepackage[varg]{txfonts}
\usepackage{graphicx,grffile, booktabs}
\usepackage{amssymb,float,natbib}
\usepackage[draft]{hyperref}
\hypersetup{colorlinks=true,citecolor=blue,linkcolor=black,
filecolor=black, pdfmenubar=true}
\bibpunct{(}{)}{;}{a}{}{,} 

\newcommand{\kms}{km s$^{-1}$}

\newcommand{\heo}{H$_2^{18}$O}

\begin{document}

\title{Volatile snowlines in embedded disks around low-mass 
protostars}

\author{D.~Harsono \inst{\ref{inst1}, \ref{inst2}, \ref{inst4}} \and 
S.~Bruderer \inst{\ref{inst3}} \and 
E.~F.~van Dishoeck \inst{\ref{inst1},\ref{inst3} }}

\institute{Leiden Observatory, Leiden University, Niels Bohrweg 2, 
2300 RA, Leiden, the Netherlands \email{harsono@strw.leidenuniv.nl} 
\label{inst1} 
\and
SRON Netherlands Institute for Space Research, PO Box 800, 9700 AV, 
Groningen, The Netherlands \label{inst2}
\and
Max-Planck-Institut f{\" u}r extraterretrische Physik, 
Giessenbachstrasse 1, 85748, Garching, Germany \label{inst3}
\and 
Heidelberg University, Center for Astronomy, Institute of Theoretical 
Astrophysics, Albert-Ueberle-Stra{\ss}e 2, 69120 Heidelberg, Germany 
\label{inst4}
}

\abstract{Models of the young solar nebula assume a hot initial disk 
such that most volatiles are in the gas phase.  Water emission 
arising from within 50 AU radius has been detected around low-mass 
embedded young stellar objects.  The question remains whether an 
actively accreting disk is warm enough to have gas-phase water up to 
50 AU radius.  No detailed studies have yet been performed on the 
extent of snowlines in an accreting disk embedded in a dense envelope 
(Stage 0). }
{Quantify the location of gas-phase volatiles in the inner envelope 
and disk system for an actively accreting embedded disk. } 
{Two-dimensional physical and radiative transfer models have been 
used to calculate the temperature structure of embedded protostellar 
systems.  The heating due to viscous accretion is added through the 
diffusion approximation.  Gas and ice abundances of H$_2$O, CO$_2$, 
and CO are calculated using the density dependent thermal desorption 
formulation. }
{The midplane water snowline increases from 3 to $\sim 55$ AU for 
accretion rates through the disk onto the star between 
$10^{-9}$--$10^{-4} \ M_{\odot} \ {\rm yr^{-1}}$.  CO$_2$ can remain 
in the solid phase within the disk for $\dot{M} \leq 10^{-5} \ 
M_{\odot} \ {\rm yr^{-1}}$ down to $\sim 20$ AU.  Most of the CO is 
in the gas phase within an actively accreting disk independent of 
disk properties and accretion rate.  The predicted optically thin 
water isotopolog emission is consistent with the detected \heo\ 
emission toward the Stage 0 embedded young stellar objects, 
originating from both the disk and the warm inner envelope (hot 
core).  An accreting embedded disk can only account for water 
emission arising from $R < 50$ AU, however, and the extent 
rapidly decreases for $\dot{M} \leq 10^{-5} \ M_{\odot} \ 
{\rm yr^{-1}}$.  Thus, the radial extent of the emission can be 
measured with future ALMA observations and compared to this 50 
AU limit. }
{Volatiles such as H$_2$O, CO$_2$, CO, and the associated 
complex organics sublimate out to 50 AU in the midplane of young 
disks and, thus, can reset the chemical content inherited from the 
envelope in periods of high accretion rates ($>10^{-5} \ M_{\odot} 
\ {\rm yr^{-1}}$).  A hot young solar nebula out to 30 AU can only 
have occurred during the deeply embedded Stage 0, not during 
the T Tauri phase of our early solar system.}

\keywords{stars: formation; accretion, accretion disks;  astrochemistry; 
ISM: molecules; stars: low-mass, stars:protostars}

\titlerunning{Volatile snowlines in embedded disks}
\authorrunning{D.~Harsono et al.}

\maketitle


\section{Introduction}\label{sec:intro}

The snowlines of various volatiles (sublimation temperature $T_{\rm 
sub} \lesssim 160$ K) play a major role for planet formation.  Beyond 
the snowline, the high abundances of solids allow for efficient 
sticking to form larger bodies, which is further enhanced by the 
presence of ices \citep[e.g.][]{stevenson88, ros13}.  Extensive
studies have investigated the snowline in protoplanetary disks around
pre-main-sequence stars similar to the nebula out of which supposedly
the solar system formed \citep[e.g.][]{lissauer87, pollack96}.  In
such models, the water snowline is located at a few AU radius.  It is
thought that the early pre-solar nebula was hot ($>1500$ K) such that
both volatiles and refractories ($T_{\rm sub} \gtrsim 1400$ K) are in
the gas phase out to larger distances \citep{cassen01, scott07,
 davis14, marboeuf14a}.  The evidence of such a hot solar nebula
comes from the history of the refractories, however the volatile
content of comets seems to indicate that a part of the disk remains
cold \citep[][]{bockelee00, mumma11, pontoppidan14}.  The evolution of
the snowline due to disk and star evolution and its accretion rate 
clearly affects the chemical composition in the region relevant to
planet formation \citep[e.g.,][]{lodders04, davis05, oberg11b}.  The
most relevant volatiles are the known major ice species: H$_2$O,
CO$_2$, and CO.  Observations \citep{meijerink09, kzhang13} 
and models \citep[e.g.,][]{dalessio98, dullemond07} of 
protoplanetary disks around pre-main sequence T-Tauri stars indicate 
that such disks are not warm enough to have gas-phase volatiles in 
the midplane beyond 30 AU as claimed in some early solar nebula 
models, and, for the case of H$_2$O, a snowline of only a few AU is 
typically found.  Higher temperatures at large radii could potentially be 
achieved, however, during the deeply embedded phase of star formation 
when the accretion rate is high.  The question remains how hot can an 
embedded accreting disk be when the accretion rate is high ($\ge 
10^{-6} \ M_{\odot} \ {\rm yr^{-1}}$, see \citealt{dunham14} for a 
recent review).

Significant progress has been made in identifying snowlines in
  protoplanetary disks in the later stages when the envelope has
  dissipated and on their location with respect to gas giant formation
  sites \citep[e.g.,][]{kennedy08, pontoppidan14}.  Direct 
  observational evidence of snowlines of the major ice species toward
  protoplanetary disks around pre-main sequence stars rely on the
  chemical changes that occur when a molecule is absent in the gas
  phase.  The most readily observed snowline is that of CO as inferred
  through spatially resolved observations of N$_2$H$^{+}$, whose 
  gas-phase abundance is enhanced when CO is frozen-out 
  \citep[e.g.][]{qi13}.  DCO$^+$ is also a tracer of cold gas at 
  temperatures close to that of the CO snowline 
  \citep{vDishoeck03, guilloteau06, qi08, mathews13}.  Both
  tracers indicate CO snowlines at $>$30 AU for T Tauri disks to
  $>$100 AU for disks around Herbig stars. The water snowline has been
  inferred to be within a few AU from direct observation and modelling
  of mid-infrared water lines \citep{meijerink09, kzhang13}.  The CO
  snowline location with respect to those of water and CO$_2$ has a
  direct impact on the amount of water present in giant planets
  atmospheres \citep{oberg11b, madhusudhan11, moses13}.

Snowlines in the early embedded stages of star formation can be
 much further out, however, since the stellar accretion process
 through the disk onto the star begins at the time that the disk
 itself is forming.  The gaseous volatile reservoir in embedded
disks is affected by their formation process, which results in
emerging protoplanetary disks having different chemical structures
depending on initial cloud core parameters \citep{visser11}.  A
related question centers on whether these volatiles are `inherited' or
`reset' during the planet formation process \citep{pontoppidan14}.
The `reset' scenario refers to the chemical processing of ices as the
gas and dust are exposed to elevated temperatures ($> 40$ K) during
their voyage from the envelope to the disk.  These temperature regions
define the regions where both CO$_2$ and CO sublimate from the ice in
the gas phase.  Both species are however difficult to trace in
  embedded disks: CO$_2$ because it lacks a dipole moment and CO
  because of confusion with the surrounding envelope.

The effect of accretion may be most readily seen through the
  changes in the water snowline.  Water is the major constituent of
  ices on the grains that facilitate planet formation \citep{gibb04,
    oberg11a} and a major coolant \citep{karska13}, and thus a key
  volatile in star- and planet formation.  Most of the water is
  thought to be formed during the pre-stellar stage, and then
  transported through the envelope into the disk and planets
  \citep{visser09, cleeves14,vDishoeck14}.  The crucial step that is
  yet relatively unexplored is the processing of water during disk
  formation.  

The water vapor content around protostars is investigated by the
`Water In Star-forming regions with Herschel' key program
\citep[WISH,][]{wish}.  Due to the large beam of {\it Herschel}, a
significant fraction of the detected water emission is from the
large-scale envelope and the bipolar outflow \citep[][]{kristensen12,
  herczeg12, mottram14}.  More importantly, the outflowing water will
escape the system and will not be retained by the disk.  In order to
determine the amount of water vapor associated with the inner envelope
or disk, an isotopolog of warm water (\heo) needs to be observed.
\citet{visser13} reported a detection of the \heo\ line (1096 GHz
$E_{\rm u}$ = 249 K)) with HIFI \citep{hifi} toward the embedded
protostar NGC 1333 IRAS2A.  However, they found that the line is still
optically thick with an emitting region of $\sim 100$ AU. Thus, it
remains difficult to constrain the amount of water vapor in embedded
disks through single-dish observations.

Spatially resolved warm \heo\ emission (excitation temperature $T_{\rm
  ex} \sim 120$ K \citealt{persson14}) has recently been detected from
within the inner 50 AU radius of several deeply embedded (Class 0)
low-mass protostars \citep[e.g.,][]{jorgensen10a, persson12,
  persson13} and toward one high mass disk \citep{vdtak06,
  kswang12}. The low-mass sources are very young objects whose
envelope mass is substantially higher than the mass at small scales
($R \la 100$ AU) \citep[also denoted as Stage 0,][]{robitaille06}.
The emission is expected to arise from $T_{\rm dust} > 100$ K regions
where ice sublimates and the gas-phase water abundance is at its
maximum \citep{fraser01, aikawa08, mottram13}.  This inner region is,
however, also where the disk forms \citep{larson03, wc11, zyli14}.
Rotationally supported disks have been detected recently around a few
low-mass protostars \citep[e.g.,][]{tobin12, murillo13b, ohashi14}.
From the kinematical information, both \citet{jorgensen10a} and
\citet{persson12} found that the water emission does not show
Keplerian motion and concluded that it must be emitted from a
flattened disk-like structure that is still dominated by the radial
infalling motions.

This paper investigates snowlines of volatiles within an accreting
disk embedded in a massive envelope.  The spatial extent and the water
vapor emission are compared with the observed values toward three
deeply embedded low-mass protostars.  The thermal structure of an
actively accreting disk is computed including the additional heating
due to the energy released from the viscous dissipation.  Most
previous studies of the thermal structure of an accreting disk focused
on the later evolutionary stage of disk evolution where the envelope
has largely dissipated away.  Furthermore, they focused on the
midplane temperature structure \citep[e.g.,][]{sasselov00, lecar06,
  kennedy08}.  The additional heating in the embedded phase shifts the
snowlines of volatiles outward to larger radii than in disks around
pre-main sequence stars \citep{davis05, garaud07, min11}.  The details
of the physical and chemical structure of the embedded disk are
presented in Section~\ref{sec:method}.  Section~\ref{sec:results}
presents the snowlines location as function of disk and stellar
properties. The results are compared with observations and their
implications on the young solar nebula is discussed in
Section~\ref{sec:dis}.  Section~\ref{sec:sum} summarizes the main
results and conclusions.


\section{Physical and chemical structures} \label{sec:method}


\begin{table}[htpb]
\centering
\caption{Parameters for the embedded disk + envelope models.  The 
varied parameters and the canonical values are indicated in boldface. }
\label{tbl:params}
\begin{tabular}{llc} \toprule \hline
Variable [unit] & Description & Value(s) \\
\hline
$r_{\rm out}$ [AU] & Outer radius & 10$^{4}$ \\
$r_{\rm in}$ [AU] & Inner radius & 0.1  \\
$\mathbf{R_{\rm cen}}$ [AU] & {\bf Centrifugal radius} & 50, {\bf 200} \\
$M_{\rm env}$ [$M_{\odot}$] & Envelope mass & 1.0 \\
$\mathbf{M_{\rm disk}}$ [$M_{\odot}$] & {\bf Disk mass} & 0.05, {\bf 0.1}, 
0.2, 0.5 \\
$\mathbf{R_{\rm disk}}$ [AU] & {\bf Disk radii} & 50, {\bf 100}, 200 \\
$H_{0}$ [AU] & Scale height at 1 AU & 0.2 \\
$T_{\star}$ [K] & Stellar temperature & 4000 \\
$M_{\star}$ [$M_{\odot}$] & Stellar mass & 0.5 \\
$\mathbf{L_{\star}}$ [$L_{\odot}$] & {\bf Stellar luminosity} & {\bf 1}, 5, 
15 \\
$\mathbf{\dot{M}}$ [$M_{\odot} \ {\rm yr^{-1}}$] & {\bf Accretion 
rate} & $10$\mbox{\small $^{ -4, -4.2, -4.5}$}$^{, {\bf -5}}$
 \mbox{\small $^{, -6, -7, -9 }$} \\

\hline
\end{tabular}
\end{table}


\subsection{Physical structure}\label{sec:phstruc}

A parametrized embedded disk (disk + flattened envelope) model is 
used to construct the density structure following \citet{crapsi08}.  
The main parameters are disk mass ($M_{\rm disk}$) and disk radius 
($R_{\rm disk}$).  A number of parameters defining the envelope and 
the disk are fixed and summarized in Table~\ref{tbl:params}.  The 
envelope mass is fixed at 1 $M_{\odot}$, which is appropriate for the 
objects from which the water emission has been detected 
\citep[][ and see Section 4.3]{prosac09, kristensen12}.  The mass 
distribution within a flattened envelope is more crucial for the 
temperature structure than the total mass.

For the large-scale envelope, a flattened envelope due to rotation as 
described by \citet{ulrich76} is adopted whose densities are given by 
the following equation: 
\begin{equation}
 \rho_{\rm env} \left (r, \mu \right) \propto \left ( \frac{R_{\rm 
cen}}{r} \right )^{1.5} \left ( 1 + \frac{\mu}{\mu_0} \right )^{-1/2} 
\left ( \frac{\mu}{2 \mu_0} + \frac{R_{\rm cen}}{r}\mu_0^2 \right 
)^{-1}, 
\end{equation}
where $\mu \equiv \cos \theta$, $R_{\rm cen}$ the centrifugal 
radius, and $r$ the spherical radius.  The centrifugal radius 
defines the region in which the material no longer flows radially and 
enters the disk.  Two centrifugal radii of $R_{\rm cen} = 50$ AU and 
200 AU are explored.  The low $R_{\rm cen}$ corresponds to the small 
Keplerian disk toward L1527 \citep{ohashi14} while the high $R_{\rm 
cen}$ corresponds to corresponds to the maximum disk radius ($\sim 
180$ AU) observed toward a Class 0 embedded low-mass YSO 
\citep{murillo13b}.  The two cases explore the effect of mass 
concentration in the inner envelope on water emission.  For a given 
centrifugal radius, a particle follows a parabolic motion given by 
\begin{equation} 
\frac{r}{R_{\rm cen}} \frac{1- \mu/\mu_0}{1-\mu_0^2} = 1,
\end{equation}
where $\mu_0$ satisfies the condition above at every $r$ and $\mu$.  
The outer radius of the envelope is fixed at $r_{\rm out} =$ 10$^4$ 
AU with an inner radius of 0.1 AU where the dust typically sublimates 
(assuming a dust sublimation temperature, $T_{\rm sub}$, between 
1500--2000 K).

An outflow cavity is then carved out from the envelope density 
structure at $\mu_0 > 0.95$.  Following \citet{crapsi08}, the density 
inside the cavity is equal to that of the densities at $r_{\rm out}$. 
This creates a conical outflow with an aperture of 30$^{\circ}$ at 
large radii (semi-aperture of 15$^{\circ}$).  For a 1 $M_{\odot}$ 
envelope, gas densities of $\sim 10^{4}$ cm$^{-3}$ fill the envelope 
cavity, which is consistent with those observed toward YSOs 
\citep[e.g.,][]{bachiller99, whitney03a}.

A flared accretion disk is added to the envelope density structure.  
The density within the disk follows the power law dependence radially 
and has a Gaussian distribution vertically in $z$ as expected from a 
hydrostatic disk.  The flared disk densities \citep{ss73, pringle81, 
hartmann98, wc11} are described by 
\begin{equation}
\rho_{\rm disk} \left ( R, z \right ) = \frac{\Sigma_0 \times 
(R/R_{\rm disk} 
)^{-1} }{\sqrt{2 \pi} H(R)} \exp \left [ -\frac{1}{2} \left 
(\frac{z}{H(R)} \right )^2 \right ],
\end{equation} 
where the scale height $H$ is fixed to 0.2 AU ($H_0$) at 1 AU 
($R_0$), $R_{\rm disk}$ the disk radius, and $R$ the cylindrical 
radius.  The radial dependence of 
the scale height is $H (R) = R \ H_0/R_0 \ (R/R_0)^{2/7}$ 
\citep{chiang97}.  Finally, the densities are scaled by a constant 
factor $\Sigma_0$ such that the total disk mass within 
$R_{\rm disk}$ ($R < R_{\rm disk}$) is equal to the values in 
Table~\ref{tbl:params} and distributed within $R_{\rm disk}$.  In 
the case of $R_{\rm cen} = 50$ AU flattened envelopes, only 
$R_{\rm disk} = $ 50 AU models are considered.  The 
total gas density in the model is $\rho = \rho_{\rm disk} + \rho_{\rm 
env}$ with a gas-to-dust mass ratio of 100.


\subsection{Temperature structure and heating terms} 
\label{sec:tempstruct}

\begin{figure}[htpb]
\centering
\includegraphics{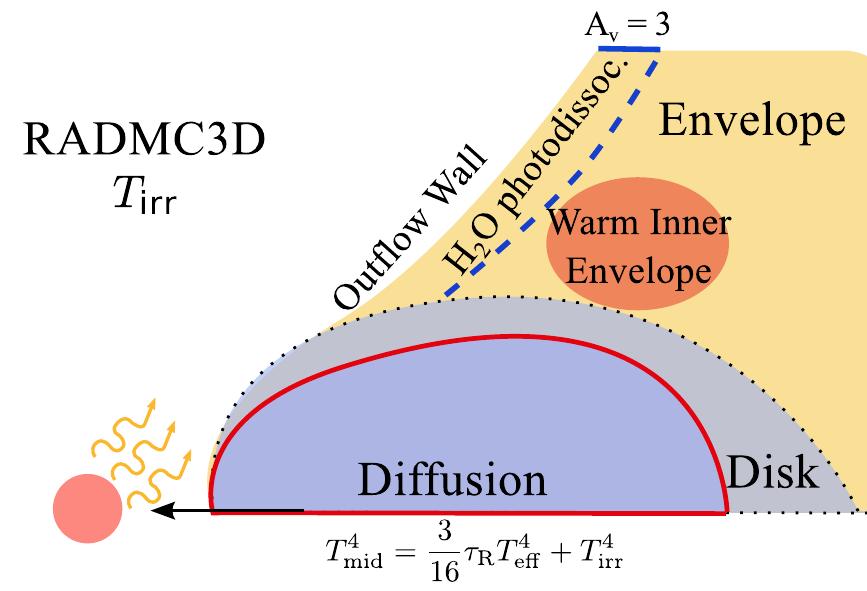}
\caption{Schematic showing the calculation of the thermal structure 
of the disk by combining both the RADMC3D Monte Carlo simulation and 
the diffusion equation.  The Monte Carlo simulation calculates the 
temperature structure due to the irradiation ($T_{\rm irr}$) from a 
central star while the diffusion equation is used to solve the 
temperature structure at high optical depths ($\tau_{\rm R} > 1$ 
where $\tau_{\rm R}$  is the Rosseland mean optical depth) as 
indicated by the red line.  This high optical depth region starts 
typically below the disk surface.  The outflow wall, warm inner 
envelope ($T_{\rm dust} > 100$ K and $r < 500$ AU) and 
photodissociation region are indicated.}
\label{fig:cart1}
\end{figure}

\begin{figure}[htpb]
\centering
\includegraphics{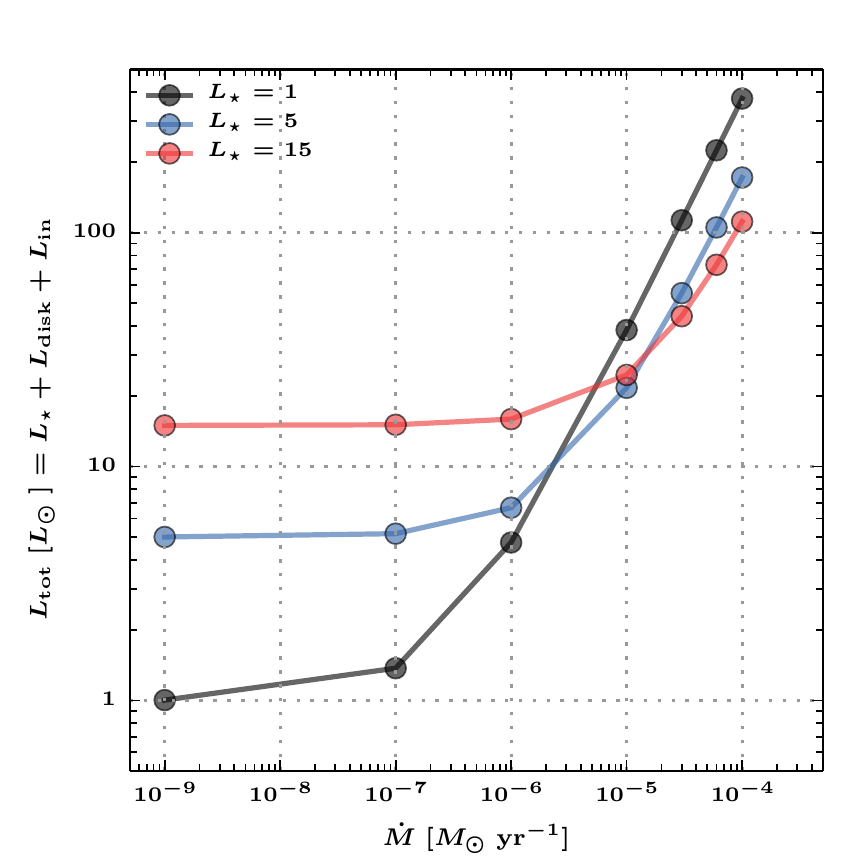}
\caption{Total luminosity ($L_{\rm tot} = L_{\star} + L_{\rm disk} + 
L_{\rm in}$) as function of accretion rates for different stellar 
luminosities.}
\label{fig:ltotmdot}
\end{figure}


The three-dimensional dust continuum radiative transfer code, 
RADMC-3D\footnote{\url{
http://www.ita.uni-heidelberg.de/\textasciitilde 
dullemond/software/radmc-3d}} is used to calculate the dust 
temperature structure.  A central star with temperature of 4000 K 
\citep[a typical inferred effective temperature of Class I protostars;
][]{white04, nisini05} characterized by $L_{\star}$ 1, 5 and 
15 $L_{\odot}$ is adopted.  An accurate dust temperature 
structure of the disk is crucial in determining the location where 
the various volatiles thermally desorb from the grain.  This is 
computationally challenging for a massive optically thick disk being 
modelled here (see \citealt{min09}).  Thus, we have separated the 
dust temperature calculations due to the central star irradiation 
(passive) from the viscous heating treatment (see 
Fig.~\ref{fig:cart1}).  The former is determined by RADMC3D 
considering a black body central star as listed in 
Table~\ref{tbl:params}. 

An actively accreting disk provides additional heating ($L_{\rm 
disk}$) from the loss of mechanical energy as the gas is viscously 
transported inward.  The steady state accretion rate is typically 
between $10^{-5} - 10^{-7} \ M_{\odot} \ {\rm yr^{-1}}$ 
\citep{hueso05}.  However, episodic accretion events such as those 
simulated by \citet{vorobyov09b} can have transient spikes with an 
accretion rate up to $10^{-4} \ M_{\odot} \ {\rm yr^{-1}}$.  Thus, 
stellar accretion rates between $10^{-8}$--$10^{-4} \ M_{\odot} \ 
{\rm yr^{-1}}$ are adopted following the $\alpha$ disk formalism 
\citep{ss73, wc11}.  The viscous heating rate per volume is given by 
\begin{equation}
 Q_{\rm visc} = \frac{9}{4} \rho_{\rm disk} v \Omega^2,
\end{equation}
where $v=\alpha c_{\rm s} H$ is the $\alpha$-dependent turbulent 
viscosity parameter, $c_{\rm s}$ the sound speed of the gas, and 
$\Omega$ the Keplerian angular velocity.  At steady state, the 
viscosity is related to the disk mass and the accretion rate through 
\citep{lodato08}:
\begin{equation}
 \dot{M} = 3 \pi v \Sigma.
\end{equation}
To explore the degree of heating, the viscosity term $v$ is varied 
for the explored accretion rates at a fixed disk mass 
($\propto \Sigma$).  The effective temperature ($T_{\rm visc}$) 
associated with the energy released at the inner radius assuming a 
hydrostatic disk is 
\begin{equation}
 \sigma_{\rm SB} T_{\rm visc}^4 \left ( R \right ) = \int Q_{\rm 
visc} \left ( R \right ) dz = \frac{3}{8\pi} \frac{GM_{\star}}{R^{3}} 
\dot{M} \left ( 1 - \sqrt{\frac{R_{\star}}{R}} \right ). 
\label{eq:tvisc}
\end{equation}
This is obtained by integrating the viscous heating terms vertically 
at all radii.  Furthermore, $T_{\rm visc}$ is the effective 
temperature of the disk at the optically thin photosphere without the 
addition of stellar irradiation.  However, the midplane temperature 
of an active disk is proportional to the vertical optical depth 
\citep{hubeny90}: $T_{\rm mid}^4 \sim \kappa_{\rm R} \Sigma_{\rm gas} 
T_{\rm visc}^4$ with $\kappa_{R}$ the Rosseland mean opacity which 
results in higher midplane temperatures.  Such a method is similar to 
that of \citet{kennedy08} and \citet{hueso05} in the optically thick 
regime.  Note that the heating from an accreting disk is caused by the 
dissipation of energy from both gas and dust.  To account for the 
irradiation from the accreting disk, $L_{\rm disk} = \int \pi \sigma 
T_{\rm visc}(R)^{4} R dR$ is added to the central luminosity 
$L_{\star}$ by determining its blackbody spectrum at $T_{\rm visc}$ 
at all radii of the disk.  For the case of high accretion rates, it 
is expected that the accretion proceeds onto the star since a gaseous 
disk is present within the dust sublimation radius.  The total 
luminosity from the inner disk is $L_{\rm in} = \int_{\rm R_{\rm 
in}}^{\rm R_{\star}} \pi \sigma T_{\rm visc}^4 R dR$.   Thus, the 
final irradiating central source $L_{\rm tot}$ is the combined 
heating from the central star, the dusty disk, and the inner gaseous 
disk: $L_{\rm tot} = L_{\star} + L_{\rm disk} + L_{\rm in}$.  
Figure~\ref{fig:ltotmdot} shows the relation between $L_{\rm tot}$ 
and $\dot{M}$ for different $L_{\star}$ values.

The following steps are taken to calculate the dust temperature of an 
accreting embedded disk.  
\begin{itemize}
 \item {Monte Carlo dust continuum radiative transfer is used to 
    simulate the photon propagation to determine the passively heated 
    dust temperature structure ($T_{\rm irr}$) due to total 
    luminosity ($L_{\rm tot}$).  The dust opacities of \citet{crapsi08} 
    are adopted and are composed of a distribution of ice coated 
    silicates and graphite grains.  The removal of ices from the grain at 
    $T_{\rm dust} > 100$ K does not strongly alter the dust temperature 
    structure. 
    }

  \item {Viscous heating is added to the region of the disk where 
    the Rosseland mean optical depth $\tau_{\rm R} > 1$.  This 
    is done by fixing the midplane temperatures to $T_{\rm mid} = 
    \left ( \frac{3}{16} \kappa_{\rm R} \Sigma_{\rm gas} T_{\rm 
    visc}^4 + T_{\rm irr}^{4} \right )^{1/4} $.  The vertical dust 
    temperature structure is calculated using the diffusion 
    approximation $\nabla D \nabla T^{4} = 0$ bounded by $T_{\rm 
    irr}$ at the surface as obtained from the previous step and 
    $T_{\rm mid}$ at the midplane where $D = \left ( 3 \rho_{\rm 
    dust} \kappa_{\rm R} \right)^{-1}$.  The diffusion is performed 
    only within the $\tau_{\rm R} > 1$ regions.  This calculation is 
    repeated a few times such that it converges.  The convergence is 
    obtained when there is no change in the the $\tau_{\rm R} > 1$ 
    region. 
    }

\end{itemize}
The addition of the viscous heating can increase the dust 
temperatures to $> 2000$ K while the typical dust vaporization 
temperature is $\sim 1500$ K.  Therefore, we have used an upper limit 
to the dust temperature of 1500 K.  Gas opacities in the inner disk 
are not taken into account, which will affect the exact temperature 
in that region.  This does not change the location of the snowlines 
since they are defined by dust temperatures $T_{\rm dust} \leq 160$ 
K.

The snowlines of protoplanetary disks without an envelope were 
obtained and compared with \citet{min11} to verify our approach 
(see Fig.~\ref{fig:mdotsnow}).  Using this formulation, the 
differences in predicted water snowlines are typically within 2 AU at 
low accretion rates and less at high accretion rates.


\subsection{Molecular abundances}



The aim of this paper is to calculate the 2D snowlines or 
snow-surfaces for CO, CO$_2$, and H$_2$O in embedded disks.  
The region in which these volatiles freeze-out onto grains depends 
on the temperature and density structure \citep[e.g.,][]{meijerink09}.  
At steady state, the rate at which the molecule is adsorbed on the grain 
is balanced by the thermal desorption rate.  We adopt dust number 
density $n_{\rm dust} = 10^{-12} \ n_{\rm H}$ \citep{visser09} with 
$a_{\rm dust} = 0.1 \ \mu$m as the effective grain size.  The number 
density of solids of species $X$ (see also Fig.~\ref{fig:gasice}) is 
simply given by
\begin{equation}
\frac{n_{\rm ice}}{n_{\rm gas}} = \frac{n_{\rm dust} \pi 
a_{\rm dust}^2 \left ( 3 k_{\rm B} T_{\rm gas} / m_X \right )^{1/2} 
}{\nu_1 \exp\left ( -E_{\rm b}/T_{\rm dust} \right ) \xi}
\label{eq:icefrac} 
\end{equation}
where $T_{\rm gas}$ gas temperature, $T_{\rm dust}$ dust 
temperature, and $m_{\rm X}$ is the mass of species X.  The 
first-order pre-exponential factor, $\nu_{\rm 1}$, is calculated 
from the binding energy, $E_{\rm b}$ \citep{hasegawa92, 
walsh10} 
\begin{equation}
\nu_1 = \sqrt{\frac{2 N_{\rm ss} E_{\rm b}}{\pi^2 m_{\rm X}}} \ {\rm 
s^{-1}}
\end{equation}
with the number of binding sites per surface area, $N_{\rm ss}$, is 
taken to be $8 \times 10^{14}$ cm$^{-2}$ following \citet{visser11}.  
A dimensionless factor $\xi \left ( n_{\rm ice} \right )$ is used to 
switch between zeroth order to first order desorption when the 
ice thickness is less than a monolayer.  The properties for each 
molecule are given in Table~\ref{tbl:molparams} along with the 
calculated pre-exponential factor $\nu_1$.  These binding energies 
assume pure ices.  The typical timescales for adsorption are 
$4 - 6 \times10^{3} \ {\rm year}$ at temperature of 50 K and 
number densities ($n_{\rm H_2}$) of $10^{6}$ cm$^{-3}$.  This 
implies that the steady state assumption is not valid at lower 
densities present in the large-scale envelope ($r > 1000$ AU) 
where the freeze-out timescales becomes longer than the 
lifetime of the core \citep[e.g.,][]{jorgensen05a}.  On the other 
hand, this paper focuses on the gas phase abundances at small-scales 
$r \leq 100$ AU where the number densities are $n_{\rm H_2} > 10^{6} 
\ {\rm cm^{-3}}$.  Photodesorption can be ignored at such high 
densities and within the disk, but it may be important at the disk's 
surface and along the outflow cavity wall.  Since the water vapor 
will also be rapidly photodissociated in those locations, these 
regions are not major water reservoirs (see Fig.~\ref{fig:cart1}).  
To approximate this region, water is assumed not to be present within 
a region that is characterized by $A_{\rm V} \leq 3$ or $N_{\rm H} 
\leq 6 \times 10^{21} \ {\rm cm^{-2}}$.  A more detailed gas phase 
abundance structure through a chemical network will be explored in 
the future.


\begin{table}[tpb]
\centering
\caption{Molecular parameters to calculate $\mathcal{R}_{\rm des}$ 
as tabulated in \citet{burke10}.} 
\label{tbl:molparams}
\begin{tabular}{lccc} \toprule \hline
Molecule & $\nu_{\rm 1}$\tablefootmark{a} & $E_{\rm b}$ & 
Refs. \\
 & [s$^{-1}$] & [K] &  \\
\hline
CO & $6.4\times10^{11}$ & 855 & \citealt{bisschop06} \\
CO$_2$ & $8.2\times10^{11}$ & 2400 & \citealt{galves07}\\
H$_2$O & $2.1\times10^{12}$ & 5773 & \citealt{fraser01} \\

\hline
\end{tabular}
\tablefoot{
\tablefoottext{a}{First order desorption pre-exponential factor 
calculated from the binding energies.}
}
\end{table}


\section{Results} \label{sec:results}

\subsection{Thermal structure of an actively accreting embedded disk}


\begin{figure}[htpb]
\centering
\includegraphics{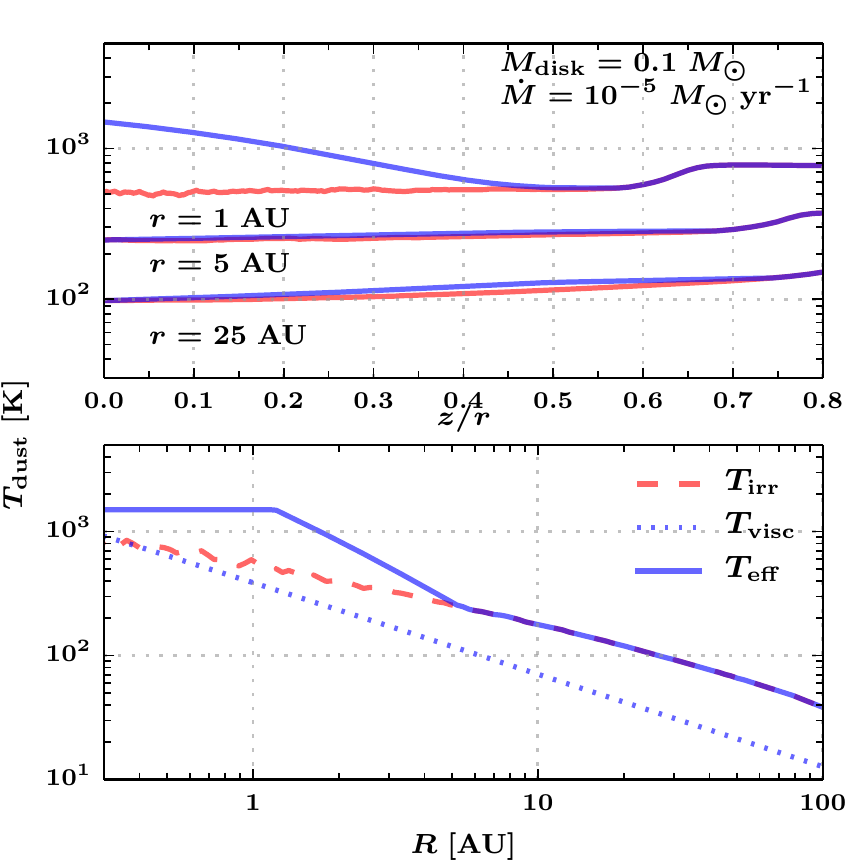}
\caption{Midplane radial ({\it bottom}) and vertical temperatures 
({\it top}) at 1, 5, and 25 AU for an embedded $0.1 \ M_{\odot}$ 
disk and an accretion rate of $10^{-5} \ M_{\odot} \ {\rm yr^{-1}}$ 
irradiated by $L_{\star} = 1 \ L_{\odot}$.  The red dashed 
({\it bottom}) and solid ({\it top}) lines indicate the thermal structure 
of a passively irradiated disk ($T_{\rm irr}$) calculated by the Monte Carlo 
simulation.  The blue solid lines indicate the temperatures including 
the viscous heating ($T_{\rm eff} = \left ( \frac{3}{16} \kappa_{\rm 
R} \Sigma_{\rm gas} T_{\rm visc}^4 + T_{\rm irr}^{4} \right )^{1/4}$). 
 The viscous temperature as calculated from Eq.~\ref{eq:tvisc} is 
indicated by the dotted blue lines.  }
\label{fig:temp1}
\end{figure}

\begin{figure*}[htpb]
\centering
\includegraphics{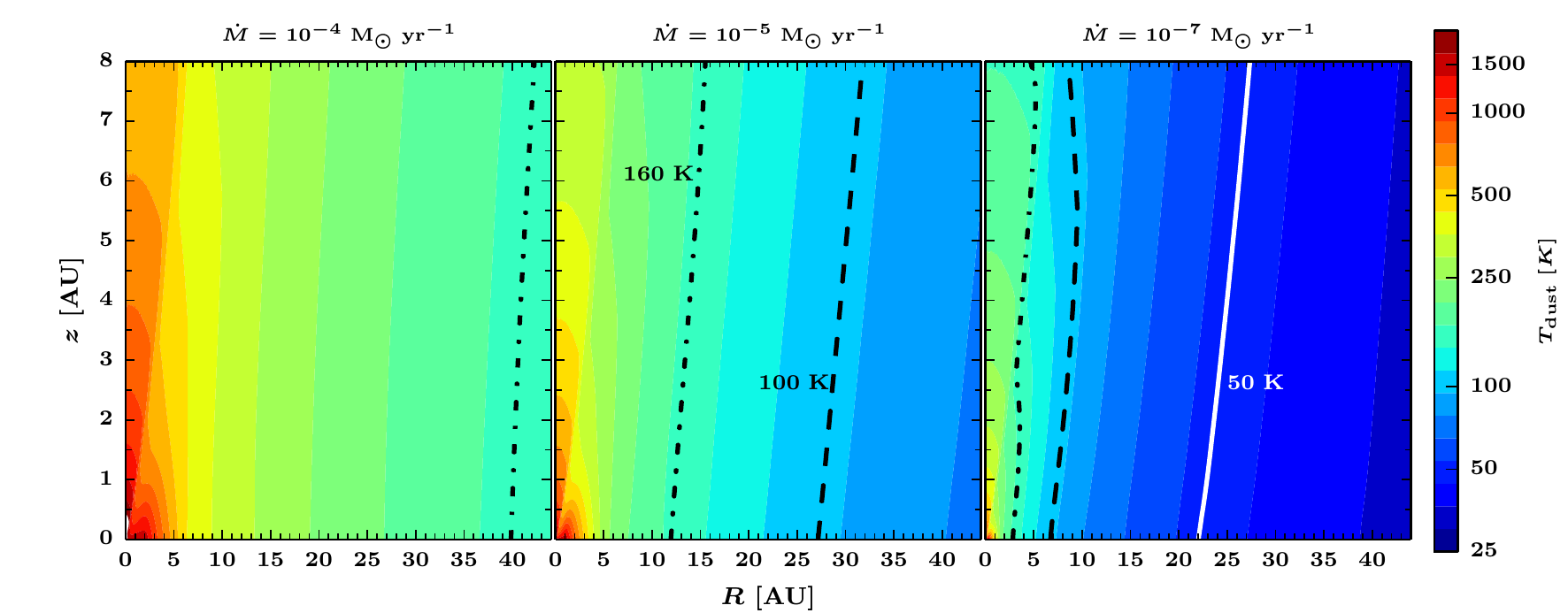}
\caption{Dust temperature structure in the inner 44 AU for three 
different accretion rates for a $0.1$ $M_{\odot}$ disk embedded in a 
1 $M_{\odot}$ envelope.  A 1 $L_{\odot}$ central heating source is 
adopted for these models.  The disk is oriented horizontally while 
the outflow cavity is oriented vertically.  The three different lines 
indicate 160 (dashed-dot) , 100 (dashed), and 50 (solid white) 
K contours, which are important for the water snowlines (see 
Fig.~\ref{fig:gasice}). }
\label{fig:temp2}
\end{figure*}

\begin{figure}[t]
\centering
\includegraphics{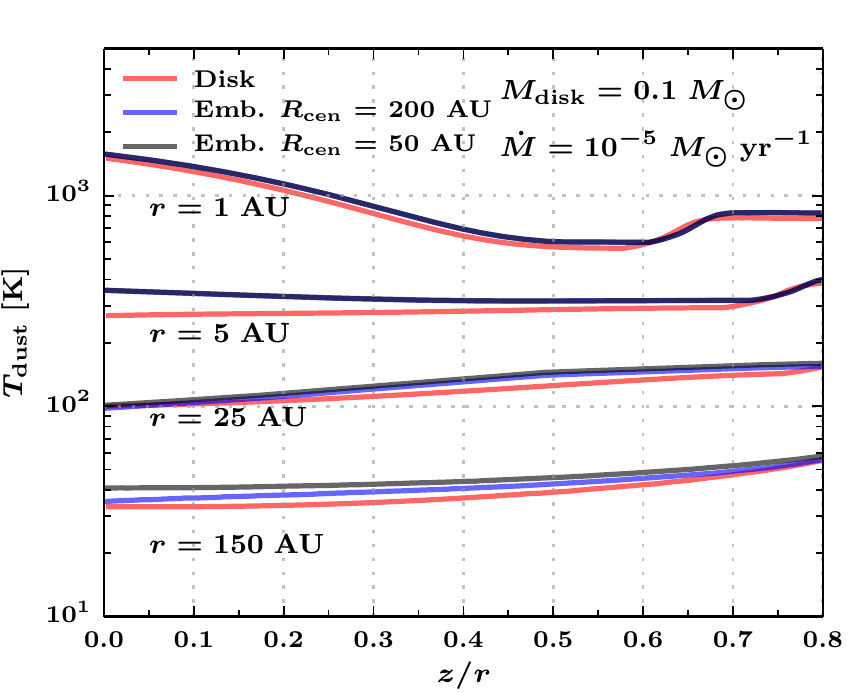}
\caption{Comparison of the vertical temperature structure at 1, 5, 
25, and 150 AU between a protoplanetary disk ({\it red}) and embedded 
disks ({\it blue} and {\it black}).  The difference between the two 
embedded disk models is the centrifugal radius $R_{\rm cen}=$ 50 AU 
({\it black}) versus 200 AU ({\it blue}). The envelope mass is $1 \ 
M_{\odot}$ for the embedded disk model.}
\label{fig:comptemp}
\end{figure}


The locations where various molecules can thermally desorb from the 
dust grains depend on the temperature structure of the disk.  
Irradiated disks have a warm upper layer with a cooler midplane.  
Figure~\ref{fig:temp1} (red dashed lines) shows the midplane ({\it 
bottom}) and vertical  temperature at a number of radii ({\it top}) 
for an embedded disk passively irradiated by a 1 $L_{\odot}$ central 
source.  Here, we present the results for a $0.1 \ M_{\odot}$ disk 
embedded in a $1 \ M_{\odot}$ envelope whose $R_{\rm cen}$ is 
200 AU (Fig.~\ref{fig:temp1} top).  These canonical parameters 
are highlighted in Table~\ref{tbl:params}.

The dust temperatures of an embedded actively accreting disk 
are indicated by the blue lines in Fig.~\ref{fig:temp1}.  The dotted line 
shows the viscous temperature ($T_{\rm visc}$) as expected at the 
photosphere (optically thin) while the solid blue lines show the effective 
dust temperature corrected for the optical depth and passive irradiation.  
As previously found, the addition of viscous heating can raise the 
temperatures in the inner few AU to $> 1000$ K 
\citep[e.g.,][]{calvet91, dalessio97, davis05}.  The disk temperature 
is above the water sublimation temperature out to $R \sim 30 $ AU for 
an accretion rate of $10^{-5} \ M_{\odot} \ {\rm yr^{-1}}$.  Due to the 
$R^{-3}$ dependence of $T_{\rm visc}$ (see Eq.~\ref{eq:tvisc}), 
the viscous heating is dominant in the inner few AU as indicated in 
Fig.~\ref{fig:temp1} ({\it bottom}).  Consequently, the passive irradiation 
due to the central luminosity ($L_{\star}$) dominates the temperatures 
along the disk's photosphere and the outflow cavity wall while the viscous 
dissipation dominates the heating deep within the disk.  These effects 
can be seen in the 2D dust temperature structure in the inner 
$\sim 40$ AU shown in Fig.~\ref{fig:temp2} for three different accretion 
rates.  For $\dot{M} \gtrsim 10^{-5} \ {\rm M_{\odot} \ yr^{-1}}$, panels 
show the vertical temperature inversions.

The dust temperature structure is also compared with a disk with 
and without an envelope whose $R_{\rm cen}$ is 50 AU in 
Fig.~\ref{fig:comptemp}. The difference is small in the inner disk, 
and lies primarily at large radii ($R > $1 AU) where the midplane 
temperature structure ($T_{\rm mid} \propto \Sigma$) is weakly 
affected by the adopted envelope model.  The temperature structure of 
the embedded disk model depends slightly on the adopted envelope 
model.  Most importantly, at $R = 5$ and 150 AU, the embedded disk 
model with $R_{\rm cen} = 50$ AU is warmer than the other two models. 
This is mainly due to the mass distribution in the inner 100 AU. 
Overall, however, the differences are small \citep[see 
also][]{dalessio97} and, for our purposes, it is sufficient to fix 
the envelope mass to 1 M$_{\odot}$ to assess the overall temperature 
structure of an embedded actively accreting disk.  The effects are 
further discussed in \S 3.4.  However, the difference at larger radii 
is sufficient to affect the dominant phase of CO$_2$ and CO.


\subsection{Water snowline}


\begin{figure}[htpb]
\centering
\includegraphics{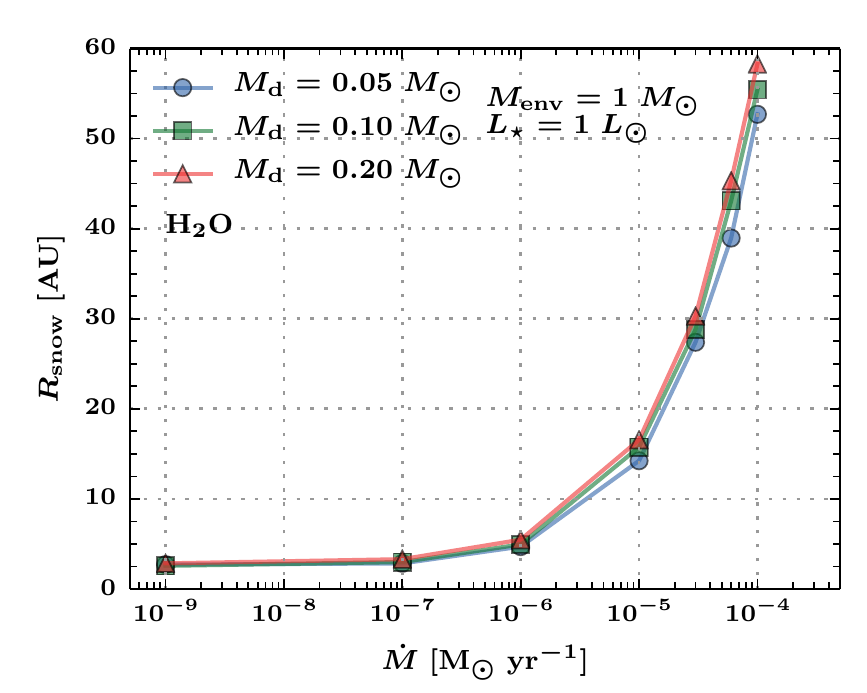}
\caption{Midplane water snowlines as function of accretion rates and 
disk mass for a 200 AU embedded disk with $R_{\rm cen}$ = 200 AU.  
The different lines indicate the snowlines dependence on disk mass 
at a fixed stellar luminosity and envelope mass. }
\label{fig:rsnow}
\end{figure}


Using the obtained dust temperatures, the gas and ice number 
densities are calculated at each cell.  To determine the snowline, 
the total available water mass was computed by adopting a water 
abundance of $10^{-4}$ with respect to H$_2$.  The total available 
mass is then multiplied by the gas fraction to determine the total 
water-vapor mass.  A minimum gas water abundance of 
$10^{-9}$ with respect to H$_2$ has been used.  The snowline 
is defined as the radius at which 50\% of the total available water 
has frozen onto the grains ($M_{\rm gas} / M_{\rm ice} = 0.5$) 
\citep[e.g.,][]{min11}.  This typically occurs at $\sim$160 K in the 
high density regions ($n_{\rm H} \sim 10^{14} \ {\rm cm^{-3}}$, see 
dashed line in Fig.~\ref{fig:temp2}).

Figure \ref{fig:rsnow} presents the midplane snowline radius as a 
function of accretion rate.  In the absence of accretion heating, the 
water snowline is located at $\sim 3$ AU.  This does not strongly 
depend on the accretion rate until a value of $\dot{M} > 10^{-7}$ 
$M_{\odot}$ ${\rm yr^{-1}}$ is reached.  The maximum water snowline 
is located at $55$ AU for an accretion rate of $10^{-4} \ M_{\odot} \ 
{\rm yr^{-1}}$.  Although the effective disk midplane temperature 
depends on the disk mass, it does not strongly affect the water 
snowline location.  It is located at only slightly smaller radius for a less 
massive disk ($0.05 \ M_{\odot}$) as indicated in Fig.~\ref{fig:rsnow}.

The steep increase of the water snowline at high accretion rates can 
be understood by comparing the stellar luminosity and the accretion 
luminosity.  For the canonical values that are indicated in 
Table~\ref{tbl:params}, the irradiating central luminosity is 1 
$L_{\odot}$.  The accretion luminosity is estimated by integrating 
Eq.~\ref{eq:tvisc} radially over the active disk between 0.1 and 200 
AU, in this case, and is approximately $L_{\rm acc} \sim 0.5 \times 
GM_{\rm star} \dot{M} / R_{\rm in}$.  Thus, the accretion luminosity 
is equal to that of the central star for $\dot{M} \sim 3 \times 
10^{-6} \ M_{\odot} \ {\rm yr^{-1}}$.  The addition of the accretion 
onto the star due to the gaseous inner disk provides an additional 
$\sim 10 \ L_{\odot}$ luminosity at the given accretion rate.  Thus, 
the accretion luminosity starts to contribute to the heating at 
$\dot{M} \gtrsim 10^{-7} \ M_{\odot} \ {\rm yr^{-1}}$ for the adopted 
parameters (see Fig.~\ref{fig:ltotmdot}).


\begin{figure}[tbp]
\centering
\includegraphics{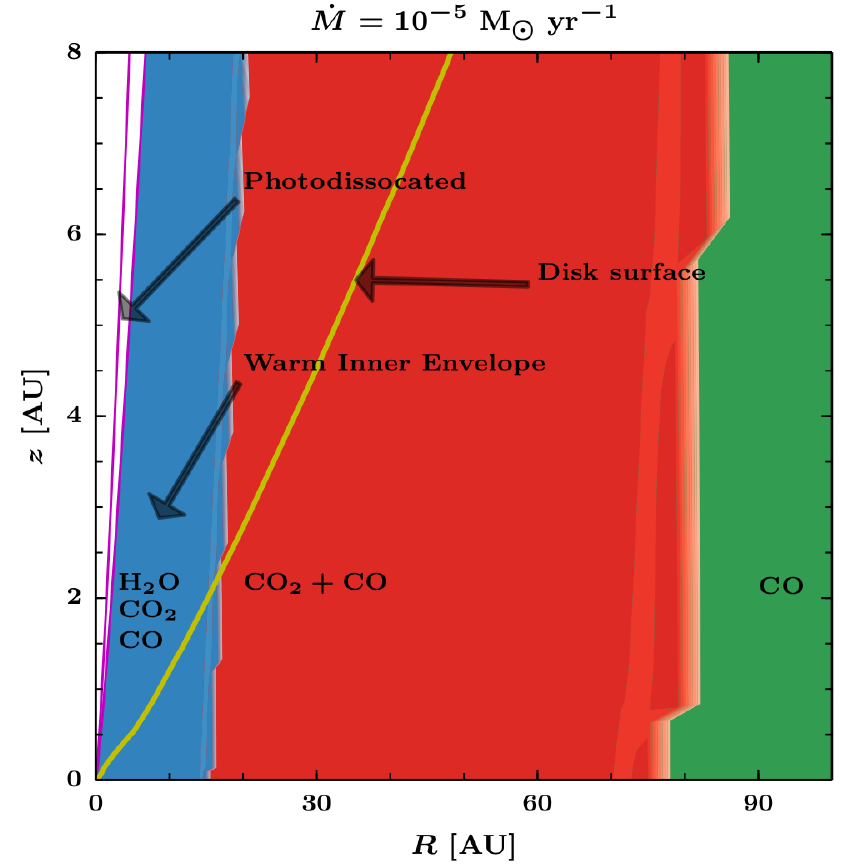}
\caption{Locations of gas phase volatiles in the 0.1 $M_{\odot}$ disk 
embedded in a 1 $M_{\odot}$ envelope.  The different colors indicate 
the different regimes where various volatiles are found in the gas 
phase: blue (H$_{2}$O, CO$_2$, and CO), red (only CO$_2$ and 
CO), and green (only CO).  The accretion rate is indicated at the 
top.  The arrows are indicating three different regions as indicated 
in Fig.~\ref{fig:cart1}: water photo-dissociation region defined by 
$A_{\rm V} = 3$, warm inner envelope, and the disk surface (yellow 
line assuming $T_{\rm gas} = T_{\rm dust}$). }
\label{fig:giratio}
\end{figure}


The observable water emission depends on the water vapor column 
density.  The water vapor column extends further than the snowline due
to the vertical gradient in water vapor abundance.  The available 
water is rapidly frozen out onto dust grains beyond the snowline.  To
determine whether the location of the snowline is within the disk or 
not, the hydrostatic disk surface is determined through $H = c_{\rm 
s}/\Omega_{\rm K}$ where $c_{\rm s} = \sqrt{k_{\rm B} T_{\rm mid} / 
\mu m_{\rm H}}$ is the sound speed and $\Omega_{\rm K}$ the Keplerian 
angular frequency.  This is the approximated regime where the gas 
should be in Keplerian motion.  For a 1 $L_{\odot}$ central star, 
most of the water emission arises from the warm inner envelope 
(sometimes also called the `hot core') above the hydrostatic disk 
surface as shown by the yellow line in Fig.~\ref{fig:giratio} if a 
constant water abundance of $10^{-4}$ with respect to H$_2$ is 
adopted.  The blue shaded region in Fig.~\ref{fig:giratio} 
indicates where water vapor is dominant, some water vapor ($< 50$\% 
in mass) is still present within the inner few AU of the red regions 
($\lesssim 30$ AU).  Water vapor is also present along the cavity walls 
as indicated by the two purple lines in Fig.~\ref{fig:giratio}, however 
these regions have low extinctions ($A_{\rm v} < 3$) which allows for 
the photodissociation of water.


\subsection{CO and CO$_2$ snowlines}


\begin{figure*}[htpb]
\centering
\begin{tabular}{cc}
\includegraphics{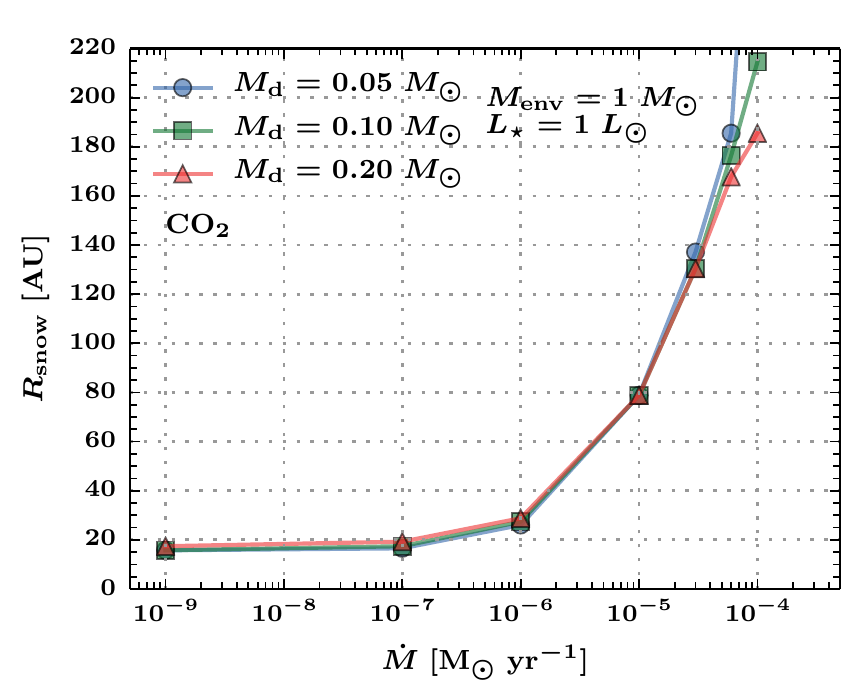}
& 
\includegraphics{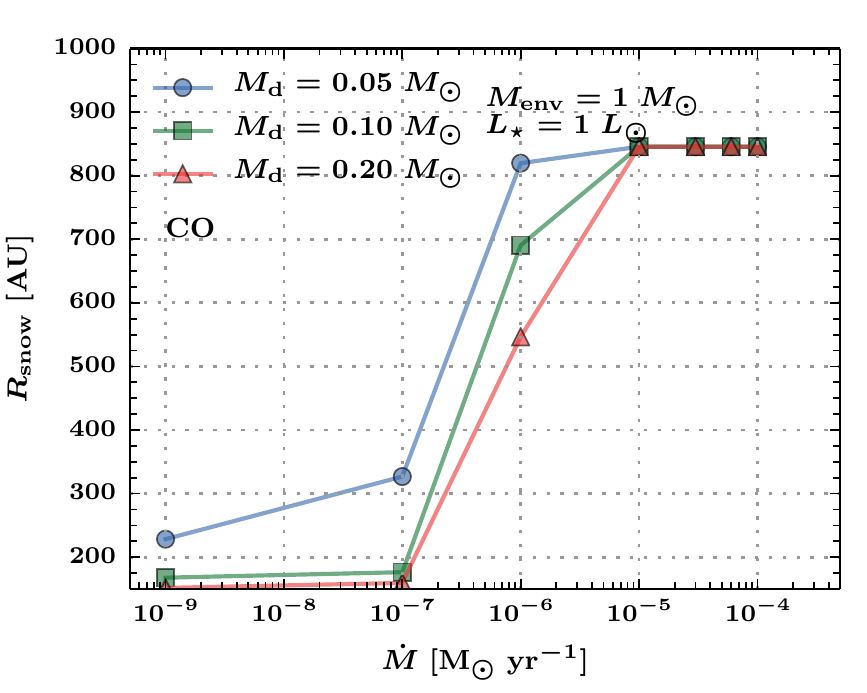}
\end{tabular}
\caption{{\it Left}: Midplane CO$_2$ snowlines as a function of 
accretion rates and disk mass for a 200 AU disk.  {\it Right}: 
Midplane CO snowlines as a function of accretion rates for the same 
disk.  The different colors indicate the snowlines dependence on disk 
mass at a fixed stellar luminosity (1 $L_{\odot}$) and envelope mass 
($1 \ M_{\odot}$).  }
\label{fig:corsnow}
\end{figure*}


Pure CO$_2$ and CO ices thermally desorb over a narrow range of dust 
temperatures between 40--80 K and 15--30 K, respectively, depending 
on the density (see Fig.~\ref{fig:gasice} in the appendix).  Due to 
their lower binding energies relative to water, CO and CO$_2$ are in 
the gas phase within a large part of the embedded disk.  The CO$_2$ 
snowline is between 20--250 AU for $10^{-9}$--$10^{-4}$ $M_{\odot} \ 
{\rm yr^{-1}}$ accretion rates (see Fig.~\ref{fig:corsnow}).  Thus, 
the entire disk including the midplane lacks CO$_2$ ice for highly 
acccreting embedded disks ($\dot{M} \sim 10^{-4} \ M_{\odot} \ {\rm 
yr^{-1}}$).  The steep rise of the CO$_2$ snowline at high accretion 
rates is similar to that of water as shown in Fig.~\ref{fig:rsnow}.

Since the adopted binding energy of CO to the dust grain is 
the least with 855 K, CO largely remains in the gas phase 
within the disk for $L_{\star} \ge 1 \ L_{\odot}$ independent of the 
accretion rate (see Fig.~\ref{fig:corsnow} right).  At low accretion 
rates, the snowline is located at $\sim150$ AU at the midplane 
indicating the presence of CO ice between 150 to 200 AU within the 
disk.  However, the bulk of CO within the disk remains in the gas 
phase as shown in Fig.~\ref{fig:giratio}.  A smaller disk during the 
embedded phase would lead to stronger envelope irradiation 
\citep[e.g.,][]{dalessio97}, and, consequently, CO is not frozen out 
within the disk.  A sufficiently large and massive disk ($M_{\rm 
disk} > 0.2 \ M_{\odot}$ irradiated by a 1 $L_{\odot}$ star) could 
contain a larger fraction of CO ice at large radii.  This is simply 
due to the increase of optical depth and, thus, lower dust 
temperatures at large radii.


\subsection{Dependence on stellar, disk, and envelope properties} 
\label{sec:34}


\begin{figure*}[tpb]
\centering
\includegraphics{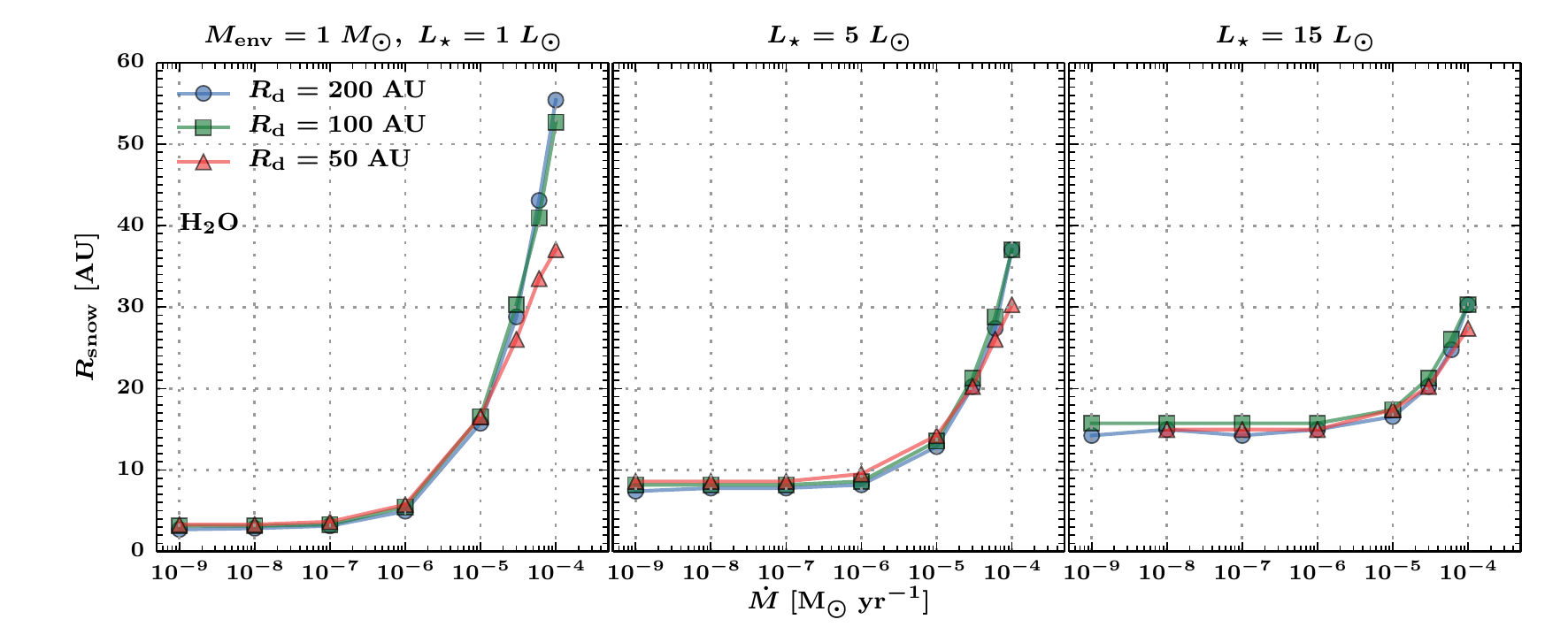}
\caption{Midplane water snowline as a function of stellar luminosity, 
accretion rate, and disk radius.  The envelope mass is fixed at 1 
$M_{\odot}$ with a disk mass of 0.1 $M_{\odot}$ and $R_{\rm 
cen}$ = 200 AU.  The different panels show the water snowline as a 
function of stellar luminosity: 1 $L_{\odot}$ ({\it left}), 5 
$L_{\odot}$ ({\it center}), and 15 $L_{\odot}$ ({\it right}).  The 
different colors indicate the disk radius: 200 AU ({\it blue}), 100 
AU ({\it green}), and 50 AU ({\it red}).}
\label{fig:giratio3}
\end{figure*}


Previous sections presented the vapor content for a $0.1\ M_{\odot}$ 
embedded disk surrounded by a 1 $M_{\odot}$ envelope with $R_{\rm 
cen}=200$~AU that is being irradiated by a 1 $L_{\odot}$ central star 
($T_{\star} = 4000$ K, $R_{\star} = 2.1 \ R_{\odot}$) as highlighted 
in Table~\ref{tbl:params}.  As noted in Section~\ref{sec:tempstruct}, 
the effective midplane temperature ($T_{\rm eff}$) depends on the 
disk mass but an increasing disk mass leads to only a small change 
to the water snowline as shown in Fig.~\ref{fig:rsnow}.  
Figure~\ref{fig:giratio} shows that the water vapor can be either 
within the disk or within the warm inner envelope.  This section 
explores how the vapor content depends on disk radius and 
central luminosity for fixed envelope and disk ($0.1 \ M_{\odot}$) 
masses.

Figure~\ref{fig:giratio3} presents the locations of water snowline as 
a function of accretion rate, luminosity, and disk radius.  For a 
fixed disk structure, a factor of 5 increase in $L_{\star}$ shifts 
the midplane water snowline by a factor of $\sim$2 from $\sim$4 AU 
(1 $L_{\odot}$) to $\sim$10 AU ( 10 $L_{\odot}$) at low 
accretion rates ($\dot{M} < 10^{-7} \ M_{\odot} \ {\rm yr^{-1}}$).  A 
factor of 15 increase in $L_{\star}$ yields a factor of $\sim 4$ increase 
in the location at the water snowline.  The stellar luminosity $L_{\star}$ 
is defined by the stellar radius while the accretion dependent total 
luminosity $L_{\rm tot}$ that is irradiating the system is shown in 
Fig.~\ref{fig:ltotmdot}.  For high accretion rates ($\dot{M} \gtrsim 
10^{-5} \ M_{\odot} \ {\rm yr^{-1}}$), the water snowline decreases 
as $L_{\star}$ increases.  This can be explained due to the fact 
that $L_{\rm tot}$ decreases as $L_{\star}$ increases at high 
accretions because $R_{\star}$ increases, which in turn decreases 
the luminosity that is provided by the inner gaseous disk.  Thus, 
if the stellar accretion rates are low ($L_{\star} > L_{\rm acc}$), the 
stellar luminosity is the parameter to be varied in order to increase 
the water snowline.  However, the water snowline can be up to 
50--60 AU if the stellar luminosity is low (small stellar radius) and  
the stellar accretion rate is high.

The CO$_2$ midplane snowline shows similar behavior as the water 
snowline (Fig.~\ref{fig:giratio3co2}) as function of stellar luminosity 
and accretion rates.  It is in the gas phase at $R > 40$ AU for 
$L_{\star} \ge 5 \ L_{\odot}$, which is a factor of 2 increase in 
terms of snowline location with respect to the canonical value 
of $L_{\star} = 1 \ L_{\odot}$.  As CO is already largely in the gas phase 
for the canonical values, a small increase of the stellar luminosity 
results in the lack of a CO snowline within the disk 
(Fig.~\ref{fig:giratio3co}).

Another parameter that affects the total energy input into the 
embedded system is the disk radius.  The total accretion luminosity 
that is being added depends on the disk radius, which is the region 
where the disk's accretion energy is being integrated.  A decreasing 
disk's radius leads to a decreasing accretion luminosity.  This only 
affects the midplane water snowline in the case of $L_{\star}$ = 1 
$L_{\odot}$ ( see Fig.~\ref{fig:giratio3} ).  In this case, the 
difference is seen only for high stellar accretion rates where the 
accretion luminosity ($L_{\rm disk} + L_{\rm in}$) starts to be the 
dominant heating source.  The midplane water snowline 
decreases from $\sim 55$ AU to $\sim 38$ AU as the disk radius 
decreases from 200 AU to 50 AU.  Thus, an increasing stellar luminosity 
($L_{\star}$) has a greater effect on the midplane water snowline.  
This leads to an increase of water vapor in the warm inner 
envelope (see Fig.~\ref{fig:giratio}) for low stellar accretion 
rates ($\dot{M} < 10^{-5} \ M_{\odot} \ {\rm yr^{-1}}$).  However, if 
$L_{\star} > 1 \ L_{\odot}$, the disk radius does not affect the 
location of the midplane water snowline even for high 
accretion rates ($\dot{M} > 10^{-5} \ M_{\odot} \ {\rm yr^{-1}}$) since 
the total luminosity is dominated by the stellar luminosity.  The 
CO$_2$ snowline, on the other hand, shows a stronger 
dependence on the disk radius than that of water.  For the high stellar 
luminosity case $L_{\star} = 15 \ L_{\odot}$, the CO$_2$ snowline 
increases from 40 AU to 80 AU as the disk radius ($R_{\rm disk}$) 
increases from 50 AU to 200 AU.  This can be explained by the 
density distribution in the inner region.  For a small disk, the dust 
temperature is low enough at large radii such that CO$_2$ can exist 
in the solid phase.  However, as the size of the disk increases, the 
overall density at similar radii decreases which results in lower 
optical depth and, as consequence, higher dust temperatures at 
large radii.  The disk radius parameter seems to be more 
important for the CO$_2$ snowline than for the water snowline.

The main parameter for the envelope in our setup is the centrifugal 
radius.  In the case of the canonical value of $R_{\rm cen} = 200$ 
AU, the mass of the disk can be distributed between 50, 100, or 200 
AU.  The disk's radius $R_{\rm disk}$ is where the disk ends and it can 
be smaller than $R_{\rm cen}$.  Here, the effects of a flattened envelope 
as defined by the centrifugal radius $R_{\rm cen}$ are presented. A 
smaller centrifugal radius results in a hotter inner disk (see 
Fig.~\ref{fig:comptemp}).  However, the affected regions correspond 
to the regions within the disk where dust temperatures are already 
$>100$ K.  Thus, we find that there is no change in the water 
snowline for a more flattened envelope structure ($R_{\rm cen} = 
50$ AU instead of 200 AU).  More volatile species such as CO$_2$ 
and CO are affected by the flattened envelope structure, however.  
The CO$_2$ snowline is confined within the inner 40 AU for 
$\dot{M} \leq 10^{-5} \ M_{\odot} \ {\rm yr^{-1}}$.  At higher 
accretion rates, the combination of a warm disk and low envelope 
densities at $R > 50$ AU for the $R_{\rm cen} = 50$ AU model results 
in a warm envelope where CO$_2$ is largely in the gas phase up to 
400 -- 500 AU along the midplane while it is $\sim 200$ 
AU in the canonical model.  In the case of CO, its snowline is 
already at 400 AU for an $R_{\rm cen} = 50$ AU envelope model at low 
accretion rates instead of $\sim 200$--300 AU in the case of $R_{\rm 
cen} = 200$ AU envelope.  Thus, we find that the inner envelope and 
disk physical structure to be important in determining the location 
of snowlines of various volatiles.  Yet, these parameters are largely 
unknown for deeply embedded young stellar objects.


\section{Discussion}\label{sec:dis}

\subsection{Comparison with accreting protoplanetary disk models}

In the absence of an envelope, similar stellar parameters and an 
accretion rate of $10^{-8} \ M_{\odot} \ {\rm yr^{-1}}$ lead to the 
midplane water snowline located at $\sim 1$ AU 
\citep[e.g.,][]{sasselov00}.  \citet{lecar06} suggest that the 
snowline can move out by increasing the accretion rate, disk mass and 
the dust opacities.  The differences between the derived snowlines 
in the literature are unlikely to be due to the differences in radiative 
transfer treatment which has been compared in \citet{min11}.  As an 
example, \citet{garaud07} derived similar water snowlines to that of 
\citet{min11} (1 AU vs 2 AU) at similar accretion rates using 
different methods in deriving the dust temperatures and different 
opacities.  The latter model takes into account the detailed 2D 
vertical structure of the disk and dust sublimation front.  As 
\citet{min11} show, most of the differences are due to the adopted 
dust opacities.  Our value is $\sim$1 AU larger than the values 
tabulated in \citet{min11} for similar parameters.  Furthermore, most 
of the previous studies adopt the minimum mass solar nebula (MMSN) 
model where $\Sigma \propto R^{-1.5}$ instead of the $R^{-1}$ 2D 
parametric disk structure used in this paper \citep[see 
e.g.,][]{andrews09}.  Previous snowlines or snow surfaces studies 
do not take the flattened envelope into account.

Previous studies have found that the water snowlines in the more
evolved disks decrease with decreasing accretion rates
\citep[e.g.,][]{davis05, garaud07, min11}.  This has been reproduced 
in Fig.~\ref{fig:mdotsnow}.  However, if the disk structure is
related to its stellar accretion rate such as adopted in
\citet{garaud07}, the water snowline increases to a constant value
at $\sim 2$ AU for very low accretion rates ($\dot{M} \lesssim
10^{-10} \ M_{\odot} \ {\rm yr^{-1}}$).  This is due to the disk
becoming more optically thin, which in turn allows for stellar
photons to penetrate deeper radially into the disk.  In this paper,
the disk structure is fixed while the stellar accretion rate is varied.  
However, the envelope provides a blanket where the disk 
can stay relatively warm. Thus, even though the disk structure 
is fixed, the snowline stays at a fixed radius for decreasing 
stellar accretion rates.

Taking the envelope into account, for the same accretion rate of 
$10^{-8} \ M_{\odot} \ {\rm yr^{-1}}$, our water snowline is at
$\sim 3$ AU for embedded disks compared to 1--2 AU without an
envelope.  The presence of the envelope does not strongly affect the
midplane temperature at regions close 100 K as shown in
Fig.~\ref{fig:comptemp}.  Thus, it does not modify the water
snowline significantly.  However it does affect the CO$_2$ and CO
snowlines since the 40 K region shifts inward under the presence of
an envelope depending on the centrifugal radius.


\subsection{Caveats}


\begin{figure*}[bpt]
\centering
\includegraphics{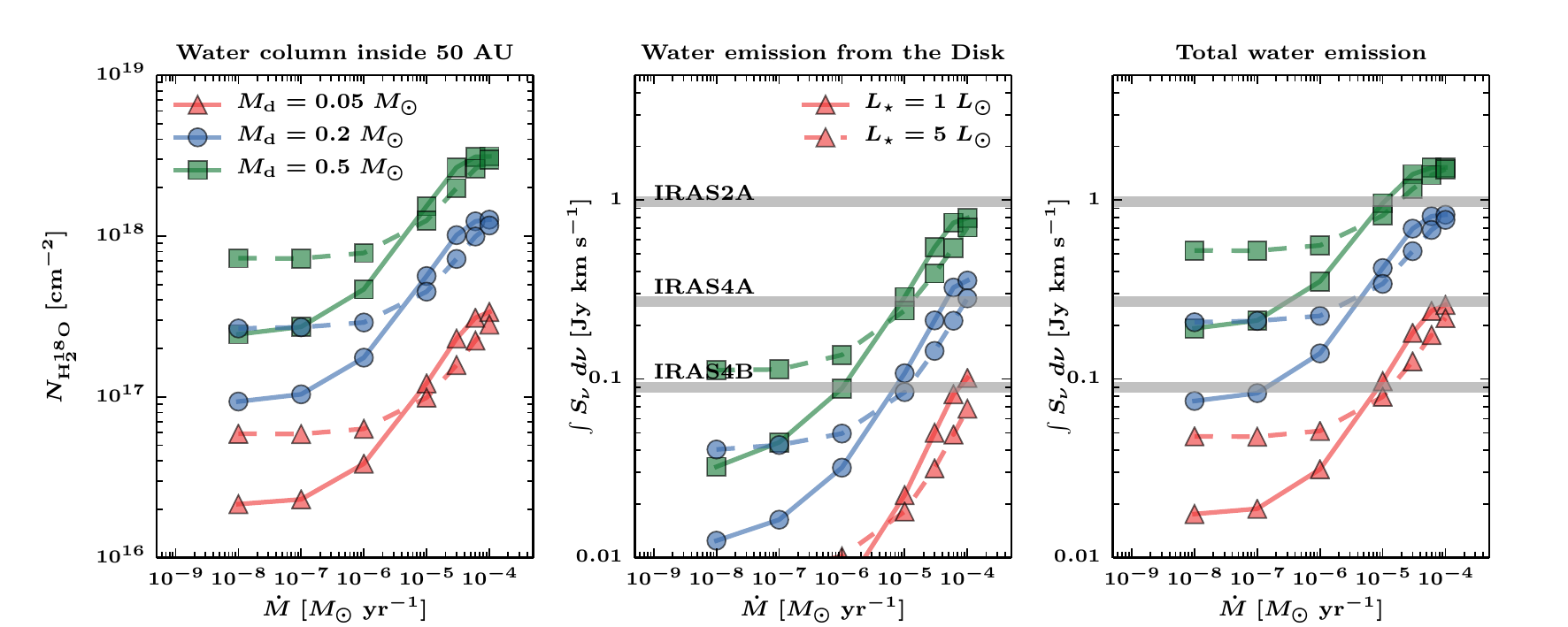}
\caption{{\it Left: } Beam averaged water (H$_2^{18}$O) column 
densities within 50 AU radius as a function of accretion rates.  The 
different colors indicate the disk mass dependence for $R_{\rm disk} 
= 200$ AU models: $M_{\rm d} = 0.05 \ M_{\odot}$ (red 
triangles), 0.2 (blue circles), and 0.5 (green squares).  The different 
lines show the luminosity dependence: $L_{\star} = 1 \ L_{\odot}$ 
(solid) and 5 $L_{\odot}$ (dashed).  {\it Center: }  Integrated line 
flux densities arising from the disk as defined by the yellow line 
in Fig.~\ref{fig:giratio}.  The observed integrated H$_2^{18}$O line 
is indicated by the gray lines for the three different embedded YSOs 
in \citet{persson14}.  {\it Right: } Integrated line flux densities 
accounting for the entire water vapor mass assuming a Gaussian 
linewidth of 1 \kms\, which is appropriate for NGC1333-IRAS4B. }
\label{fig:waterintens}
\end{figure*}


The effect of convection in the vertical direction is not included in 
this study.  It is typically found to be important in the case of 
high accretion rates ($\dot{M} > 10^{-6} \ M_{\odot} \ {\rm 
yr^{-1}}$, \citealt{dalessio98}, \citealt{min11}).  Convection is 
found to cool the midplane temperatures at $T_{\rm dust} > 500$ K.  
Thus, this should not affect the water (100--160 K), CO$_2$ ($\sim 
50$ K) and CO ($\sim 20$ K) snowlines as their sublimation 
temperatures are much lower than 500 K.

Under the assumption of the steady state accretion disk model, the 
typical values of $\alpha$ are found to be $> 1$ for the case of an 
accretion rate of $\dot{M} = 10^{-4} \ M_{\odot} \ {\rm yr^{-1}}$.  
This is significantly larger than that expected from 
magnetorotational instability (MRI) driven accretion in a MHD disk 
\citep[$\alpha = 0.01$,][]{bh98}.  \citet{kpl07} indicated that 
$\alpha$ could be $\sim 0.4$ for thin disks around compact systems.  
Such large $\alpha$ values in our models are obtained since we have 
fixed the density distribution at a high accretion rate ($\dot{M} > 
10^{-5} \ M_{\odot} \ {\rm yr^{-1}}$).  In reality, such a disk is 
unphysical.  However, this is only encountered for the highest 
accretion rate of $10^{-4} \ M_{\odot} \ {\rm yr^{-1}}$ that occurs 
for a very short time.  In principle, this can be avoided by 
iterating the disk structure to ensure that the hydrostatic 
conditions are satisfied.  However, this paper explores the effect of 
the adopted accretion rates and disk masses for a given fixed set of 
parameters.  Self-consistent models including gas opacities and 
multiple dust species should be explored in the future.


\subsection{Comparison with observations}

The extent of the observed spatially resolved water emission toward 
low-mass embedded YSOs is between 25--90 AU \citep{jorgensen10a, 
persson12, persson14}.  The water snowline is confined to within the 
inner 40 AU for the cases of the envelope models whose centrifugal 
radius is 50 AU.  In such models, we have chosen that the disk does 
not extend beyond 50 AU.  The water snowline is only increased in 
cases where the centrifugal radius of the envelope is larger than 50 
AU, in which case the water snowline can extend up to 55 AU if the 
stellar luminosity is low ($L_{\star}$ = 1 $L_{\odot}$).  Thus, our 
results indicate that most of the water emission at $R < 50$ AU would 
originate from the combined disk and warm inner envelope (see 
Fig.~\ref{fig:giratio}).

Through the combined modelling of the continuum spectral energy
  distribution and submm continuum images, the envelope masses 
  toward the observed embedded sources (NGC1333 IRAS2A, 
  IRAS4A, and IRAS4B) are found to be between 3 -- 5 $M_{\odot}$ 
  \citep{kristensen12}, somewhat larger than our canonical value.  On 
  the other hand, all three sources are binaries \citep[e.g.,][]{
  jorgensen04b, prosac07} so it is likely that the effective envelope 
  mass that affects the disk physical structure is lower than the above 
  values.  Therefore, the envelope masses have been kept fixed at 
  1 $M_{\odot}$.

The amount of water vapor in the embedded disk models can be compared 
to the observed optically thin millimeter water emission toward 
deeply embedded YSOs.  Thermalized H$_2^{18}$O emission at 203.4 GHz 
($3_{\rm 1, 3} - 2_{\rm 2,0}$) is calculated using the following 
equation 
\begin{eqnarray}
 I_{\nu} & \approx & B_{\nu} \left ( T_{\rm ex} \right ) (1 - 
e^{-\tau}), 
\\
\tau & = & \frac{\bar{N}}{\mathcal{Q} \left ( T_{\rm ex} \right ) 
\Delta V} \frac{A_{\rm ul} c^3}{8 \pi \nu^3} e^{h \nu / k_{\rm B} 
T_{\rm ex}} ,
\end{eqnarray}
where $A_{\rm ul}$ is the Einstein $A$ coefficient of the 203 GHz 
transition is $4.5\times10^{-7} \ {\rm s^{-1}}$, $T_{\rm ex}$ is 
taken to be 150 K, $E_{\rm u} = 204$ K is the upper energy level, 
and $\mathcal{Q}$ is the temperature dependent partition function 
adopted form the HITRAN database \citep{hitran}.  The beam 
averaged column density ($\bar{N}$) is taken from within 50 AU 
radius with $^{16}$O:$^{18}$O isotopic ratio of 540 \citep{wilson94}.  
The line profile is assumed to be Gaussian with a $FWHM$ ($\Delta 
\upsilon$) = 1--4 km s$^{-1}$ as observed toward the three embedded 
sources reported by \citet{persson14}.

Figure~\ref{fig:waterintens} ({\it left}) presents the mean 
H$_{2}^{18}$O gas column density of the simulated embedded disks
as a function of accretion rate for different disk masses and stellar 
luminosities.  The mean column density rises sharply for $\dot{M} > 
10^{-7} \ M_{\odot} \ {\rm yr^{-1}}$ as expected for a 1 $L_{\odot}$ stellar 
luminosity while the sharp rise occurs at $\dot{M} > 10^{-6} \ M_{\odot}$ 
for a 5 $L_{\odot}$ irradiating source.  The typical line optical depth at 
the line center ($\upsilon = 0$ \kms) can be $\tau_{\rm \upsilon = 0} 
\ge 1$.  The typical average column densities of water between 50 AU 
to 1000 AU are $>4$ orders of magnitude lower than the values 
within 50 AU as presented in Fig.~\ref{fig:waterintens}.  Thus, the 
water present within 50 AU dominates the total water mass for the 
adopted models.

The expected integrated H$_2^{18}$O 3$_{1, 3}$--2$_{2, 0}$ (203.4 
GHz) line flux densities for the embedded disk models are also shown 
in Fig.~\ref{fig:waterintens} ({\it right}).  The integrated line 
flux densities ($\int S_{\nu} d\nu$) are calculated with a Gaussian 
line profile with $\Delta v = 1$ \kms\ ($1.06 \ S_{\rm \upsilon =0} 
\times \Delta \upsilon$ where $S_{\upsilon = 0}$ is the peak flux 
density at line center) as observed toward NGC1333-IRAS4B 
\citep{persson14}.  The observed line widths toward IRAS2A and IRAS4A 
are 4 and 3 \kms, respectively.  Increasing $\Delta \upsilon$ 
slightly increases the model integrated line flux densities because 
the line is slightly optically thick.

The predicted integrated line flux densities from actively accreting 
embedded disk models are consistent with that observed toward NGC1333 
IRAS4B if $M_{\rm d} \leq 0.2 \ M_{\odot}$ with $\dot{M} \gtrsim 6 
\times 10^{-7} \ M_{\odot} \ {\rm yr^{-1}}$ for $L_{\star} = 1 \ 
L_{\odot}$.  \citet{prosac09} estimated from the continuum  that the 
disk mass is $\sim 0.2 \ M_{\odot}$, which is consistent with our 
results.  Decreasing the disk mass implies a higher accretion rate is 
required to reproduce the observed water emission.  A higher accretion 
rate ($\sim 1 \times 10^{-5} \ M_{\odot} \ {\rm yr^{-1}}$) with lower 
disk mass is also consistent with the observed extent of water 
emission at $\sim 20$ AU.  A similar conclusion is reached for the 
case of IRAS4A whose disk is estimated to be $\sim 0.5 \ M_{\odot}$ 
\citep{prosac09}.  Since the bolometric luminosities toward NGC1333 
IRAS4A and IRAS4B are a factor of 4--9 higher than the adopted 
central luminosity ($L_{\star}$), a model with a factor of 2 lower 
mass disk can also reproduce the observed flux density with 
accretion rates $\sim 10^{-5} \ M_{\odot} \ {\rm yr^{-1}}$.  The 
total luminosity generated by such embedded systems would be similar 
to the observed bolometric luminosities.  With such a lower mass 
disk, the warm inner envelope is the dominant source of emission 
instead of the Keplerian disk.  This implies that it would be 
difficult to observe the Keplerian motion of the disk from 
H$_2^{18}$O lines.

IRAS2A shows stronger integrated line emission ($\sim 1$ Jy \kms) 
relative to the other 2 sources and extending up to 90 AU radius.  As 
pointed out above, a disk cannot be responsible for any warm water 
emission at $R > 50$ AU.  Furthermore, \citet{brinch09} estimated 
that the disk is at most $\sim 0.05 \ M_{\odot}$, which lowers the 
expected disk's contribution to the observed water emission (see also 
\citealt{prosac09} and \citealt{maret14}).  There are two possible 
scenarios that can explain such a high water emission: a more 
luminous star ($L_{\star} > 10 \ L_{\odot}$) or a high 
stellar accretion rate ($\sim 10^{-5} \ M_{\odot} \ {\rm yr^{-1}}$).  
Both solutions result in total luminosity $L_{\rm tot}$ comparable to 
the observed bolometric luminosity of IRAS2A ($L_{\rm bol} = 35.7 \ 
L_{\odot}$, \citealt{karska13}).  In either case, most of the water 
emission arises from the warm inner envelope with negligible disk 
contribution.

The best cases for an actively accreting embedded disk scenario are 
IRAS4A and IRAS4B.  More detailed modelling of their physical 
structure at $<100$ AU radius is required to further constrain the 
relation between infall (envelope to disk) and stellar accretion rate 
(disk to star).  \citet{mottram13} suggests that the infall rate from 
the large-scale envelope toward IRAS4A at 1000 AU to be $\sim 10^{-4} 
\ M_{\odot} \ {\rm yr^{-1}}$.   For the cases of disk masses between 
0.2--0.5 $M_{\odot}$, the disk must also process the material at 
similar rate (within a factor of 10), which is consistent with the 
models presented here.  

\paragraph{Observational perspectives}
Future spatially and spectrally resolved ALMA observations of 
optically thin water (H$_{2}^{18}$O ) and deuterated water (HDO) lines 
may be able to differentiate these models.  The spatially resolved data 
can provide the extent of the water emission.  The comparison between 
the source's bolometric luminosity and the extent of the water emission 
can provide limits on the stellar accretion rate.  A low bolometric 
luminosity source cannot have a water emission originating from the 
disk further than a few AU.   Furthermore, the velocity information 
within the inner 50 AU radius can help in differentiating between 
the disk and the infalling envelope \citep[see e.g.,][]{harsono15a}.


\subsection{Connecting to the young solar nebula}

It is thought that the disk out of which the solar system formed (the
solar nebula) was initially hot enough to vaporize all material 
inherited from the collapsing cloud into atoms, then turn them
into solids according to a condensation sequence as the nebula cools 
\citep{lewis74, grossman74}.  Evidences for energetic processing and 
subsequent condensation comes from meteoritic data collected in the 
inner solar system \citep[see][for reviews]{kerridge88, scott07, 
apai10}.  After the refractory phases have formed, more volatile 
species such as ices can also condense at larger distances from the 
young Sun, with ice composition depending on  the temperature and 
pressure as well as the elemental abundance ratios 
\citep[e.g.,][]{lunine91, owen93, mousis09, pontoppidan14}.  Using 
cooling curves appropriate for the solar nebula, models show that 
various ices including H$_2$O, CH$_3$OH, NH$_3$ and CO$_2$ ice form 
by condensation out of warm gas out to at least 20 AU 
\citep{mousis12a, marboeuf14a}.  This condensation process is 
important to explain the composition of volatiles in the atmospheres 
of solar-system giant planets and comets.

Where and when does this heating and condensation in the disk 
actually occur?  The above processes are usually discussed in the 
context of protoplanetary disks where the envelope has dissipated and 
where the water snowline is eventually at a few AU once the disk has 
cooled.  However, neither observations nor models of disks around 
solar-mass T-Tauri stars show any evidence that disks are as warm as 
required above to have condensation happening out to 20 AU 
\citep{beckwith90, dalessio98, dullemond07}.  Furthermore, 
observations suggest that significant grain growth has occurred 
\citep{testi14} and disk inhomogeneities due to protoplanets are 
present in such evolved disks \citep[e.g.,][]{quanz13}.  This further 
violates the assumption that the physical structure of an evolved 
disk can be used to study the young solar nebula.  The results from 
this paper indicate that instead such a hot disk may be more common 
in the early stages of disk formation during the {\it embedded} phase 
($M_{\rm env} > M_{\rm disk}$).  The disk is then expected to 
have high accretion rates providing the necessary additional heating.  
Furthermore, the high accretion rates necessary to push the midplane 
snowlines to 20 AU and beyond tend to occur for only a short time 
\citep[e.g.,][]{vorobyov09b}.  Most volatiles including H$_2$O, 
CH$_3$OH, CO$_2$ and CO presented here will then be in the gas-phase 
within this 20--30 AU radius through sublimation of ices.  Shortly 
after, the volatiles re-condense as the accretion rate decreases, 
with the process controlled by the freeze-out timescale.  The 
inclusion of the gas motions (radially and vertically) and kinetics 
does not allow for instantaneous re-condensation, thus some gas-phase 
volatiles should remain near the snowline \citep{lewis80, ciesla06}.

A related question is whether the volatiles that are incorporated 
into planetesimals and eventually planets and icy bodies in the 
critical 5--30 AU zone are then {\it inherited} or {\it reset} during 
the disk formation process.  The material is defined to be {\it 
inherited} if the simple and complex ices that have been built up in 
the envelope survive the voyage to the planetary and cometary forming 
zones.  Evolutionary models by \citet{visser09} and \citet{visser11} 
suggest that strongly bound ices such as H$_2$O are largely pristine 
interstellar ices in the outer disk, whereas more volatile species 
can have sublimated, recondensed and reprocessed several times on 
their way to the inner disk.  However, these models have not included 
internal viscous dissipation as an additional heating source.  A 
warmer early disk can facilitate ice evolution and formation of 
complex organic molecules in the temperature regime between 20--40 K 
\citep{garrod08, nomura09, walsh14, drozdovskaya14}.  Our results 
including the accretion heating indicate that significant ice 
re-processing may occur out to larger radii than thought before, well 
in the comet formation zone.  Such a {\it reset} scenario is 
supported by the meteoritic data from the inner few AU, but the 
fraction of material that is reset at larger radii is still unknown 
\citep[e.g.][]{pontoppidan14}.  If some of the icy grains are 
incorporated into km-sized bodies (planetesimals) immediately 
following disk formation, a larger fraction of the  material could 
still remain pristine.


\section{Summary and conclusions}\label{sec:sum}

Two-dimensional embedded disk models have been presented to 
investigate the location of the snowlines of H$_2$O, CO$_2$ and CO.  
The dust temperature structure is calculated using the 3D dust 
radiative transfer code RADMC3D with a central heating source 
($L_{\star} + L_{\rm visc} + L_{\rm in}$).  An additional heating 
term from an actively accreting disk has been added through a 
diffusion approximation.  The main parameters that are explored in 
this paper are $L_{\star}$, centrifugal radius, disk radius, disk 
mass, and stellar accretion rate.  In addition, the extent of the 
snowline and optically thin water emission are compared with 
observations toward three deeply embedded low-mass YSOs.  The 
following lists the main results of this paper. 

\begin{itemize}

    \item The presence of an envelope serves as a blanket 
    for the disk such that the dust temperature within the inner few 
    AU stays relatively warm for the adopted disk parameters.  The 
    adopted centrifugal radius affects the temperature structure of the 
    disk since the effective midplane temperature depends on 
    the optical depth.  An envelope with a smaller centrifugal radius 
    distributes more mass in the inner 100 AU than an envelope with 
    a higher centrifugal radius.  Such high concentration of mass in the 
    inner 100 AU increases the effective Rosseland mean optical depth 
    that an observer at the disk's midplane sees vertically, which in turn 
    increases the effective midplane temperature.  While it does not 
    drastically affect the $T_{\rm dust} > 100$ K regime, the presence 
    of an envelope affects the location of $T_{\rm dust} \lesssim 40$ K, 
    which is important for the CO and CO$_2$ snowlines.

    \item  The midplane water snowline in the models can extend up 
    to a maximum of $\sim 55$ AU for a disk accreting at $10^{-4} \ 
    M_{\odot} \ {\rm yr^{-1}}$.  The CO$_2$ snowline is located at 
    $\gtrsim$100 AU for the same accretion rate.  Both H$_2$O and 
    CO$_2$ can remain in the solid phase at large radii ($R > 100$ AU) 
    in the midplane within the boundaries of an embedded hydrostatic disk.

  \item CO is largely found to be in the gas phase within the embedded
    disk independent of accretion rate and disk properties.  Some CO
    could be frozen out at large radii in the midplane but only for a
    relatively massive embedded disk ($M_{\rm d} \ge 0.1 \
    M_{\odot}$).

  \item The CO$_2$ and CO snowlines are affected by the exact
    structure of the flattened envelope.  A smaller centrifugal radius
    corresponding to a highly flattened envelope model increases the
    CO and CO$_2$ snow surfaces to larger radii.

  \item The observed H$_2^{18}$O emission toward NGC1333 IRAS4A and
    IRAS4B is consistent with models for $\dot{M} > 10^{-6} \
    M_{\odot} \ {\rm yr^{-1}}$ for the adopted disk masses of 0.5
    $M_{\odot}$ and 0.2 $M_{\odot}$, respectively.  The observed size
    of water emission can also be explained with a model with a
    slightly higher accretion rate but with a lower disk mass ($R
      < 50$ AU).  Toward IRAS2A, the models suggest a smaller disk
    contribution than toward IRAS4A and IRAS4B.  Most of the
      optically thin water emission is emitted from the warm inner
      envelope (hot core) for that source ( see
      Fig.~\ref{fig:giratio}).

  \item Significant chemical processing is expected to occur in the
    inner 100 AU region during the disk formation process in Stage 0.
    Midplane temperatures between 20--40 K are expected to be
    prevalent up to 100 AU radius, and 100 K out to 30 AU, but only
    for the highest accretion rates found in the early phases of star
    formation.  During these warm phases, the chemistry inherited from
    the collapsing cloud could be reset and more complex molecules
    could form, unless the ices are sequestered early into
    planetesimals.

\end{itemize}

In connection to early solar system formation, an important conclusion
from our models is that a hot young solar nebula with water vapor out 
to 30 AU can only occur during the deeply embedded phase, not the T 
Tauri phase of our solar system.  Our models also show that most of 
the observed optically thin water emission arises within the inner 50
AU radius.  Future Atacama Large Millimeter/submillimeter Array (ALMA)
observations spatially resolving the inner 50 AU can test our 
embedded accreting disk models when coupled with better physical 
models for individual sources.  This will lead to better 
understanding of the physical and chemical structure and evolution of 
disks in the early stages of star formation.


\section*{Acknowledgments}

We thank Atilla Juh{\'a}sz for providing the scripts for generating and 
analyzing RADMC3D input and output files.  We also thank Kees 
Dullemond for providing RADMC3D.  We are grateful to Catherine Walsh, 
Lee Hartmann, Joseph Mottram and Lars Kristensen for fruitful 
discussions.  We are also grateful to the anonymous referee whose 
comments have helped to improve the paper.   This work is supported by the 
Netherlands Research School for Astronomy (NOVA).  Astrochemistry in 
Leiden is supported by the Netherlands Research School for Astronomy 
(NOVA), by a Royal Netherlands Academy of Arts and Sciences (KNAW) 
professor prize, and by the European Union A-ERC grant 291141 
CHEMPLAN.  DH is funded by Deutsche Forschungs-gemeinschaft 
Schwerpunktprogram (DFG SPP 1385) The First 10 Million Years of the 
Solar System -- a Planetary Materials Approach.


\bibliographystyle{aa}
\bibliography{../../biblio.bib}


\appendix
\Online


\section{Snowline test}

\begin{figure}[htpb]
 \centering
 \includegraphics{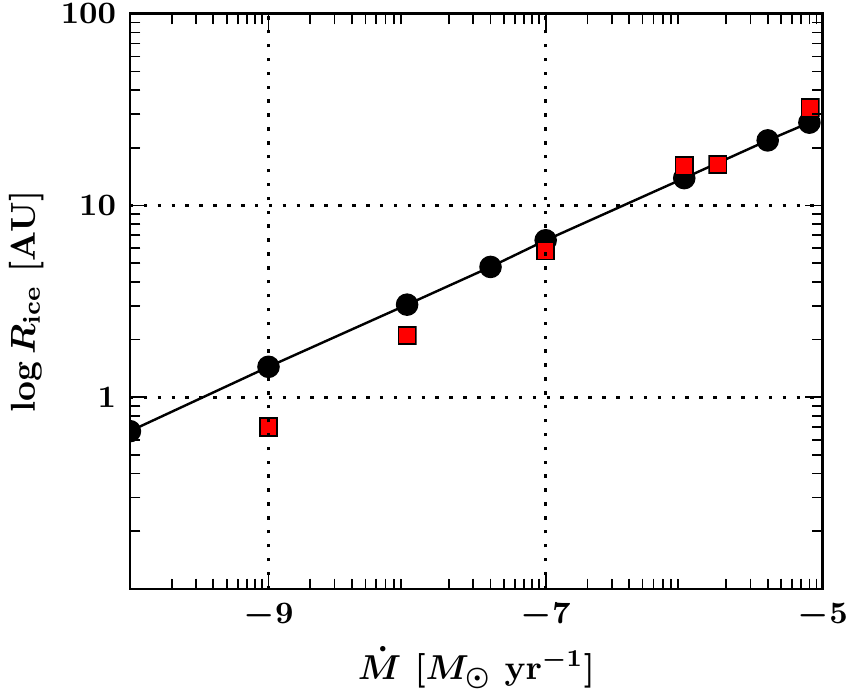}
\caption{Snowlines for the MMSN disk without an envelope: black 
circles show the radii calculated with our method and red squares are 
tabulated values from \citet{min11}.}
\label{fig:mdotsnow}
\end{figure}

Water snowlines were inferred and compared with results from 
\citet{min11} in Fig.~\ref{fig:mdotsnow} using the minimum mass solar 
nebula (MMSN) model ($\Sigma \propto r^{-1.5}$).  The comparison 
shows that our adopted method reproduces the water snowlines at 
high accretion rates.  For low accretion rates $ \dot{M} \leq 10^{-9} 
\ M_{\odot} \ {\rm yr^{-1}}$, our values are slightly smaller yet 
consistent with those reported in literature.    

\section{H$_2$O, CO$_2$, and CO gas fraction}

\begin{figure}[htpb]
\centering
\includegraphics{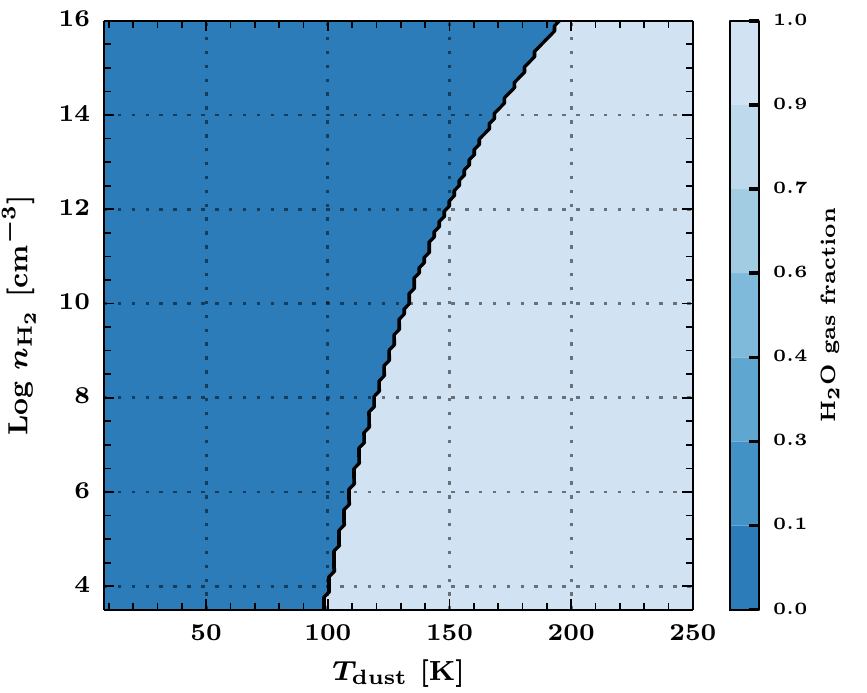}
\includegraphics{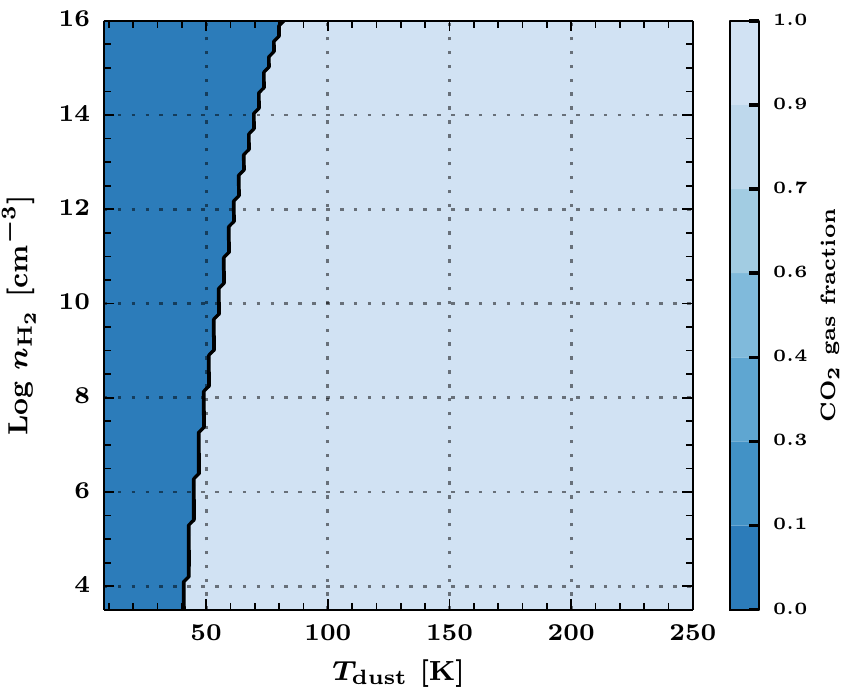}
\includegraphics{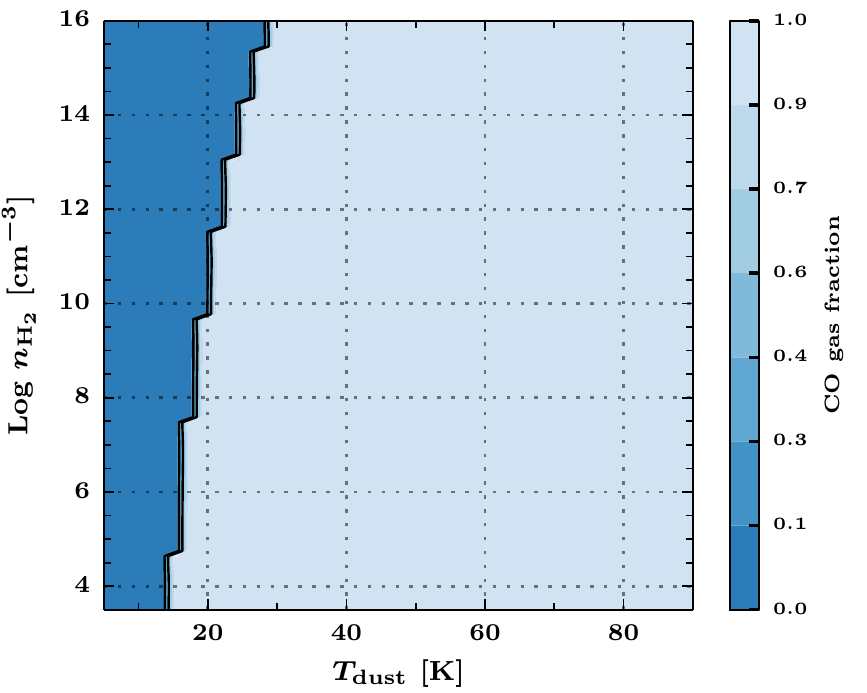}
\caption{Gas fraction ($n_{\rm gas}/n_{\rm gas} + n_{\rm ice}$) for 
H$_2$O ({\it top}), CO$_2$ ({\it middle}) and CO ({\it bottom}) as 
a function of density and temperatures.}
\label{fig:gasice}
\end{figure}

\begin{figure*}[bt]
\centering
\includegraphics{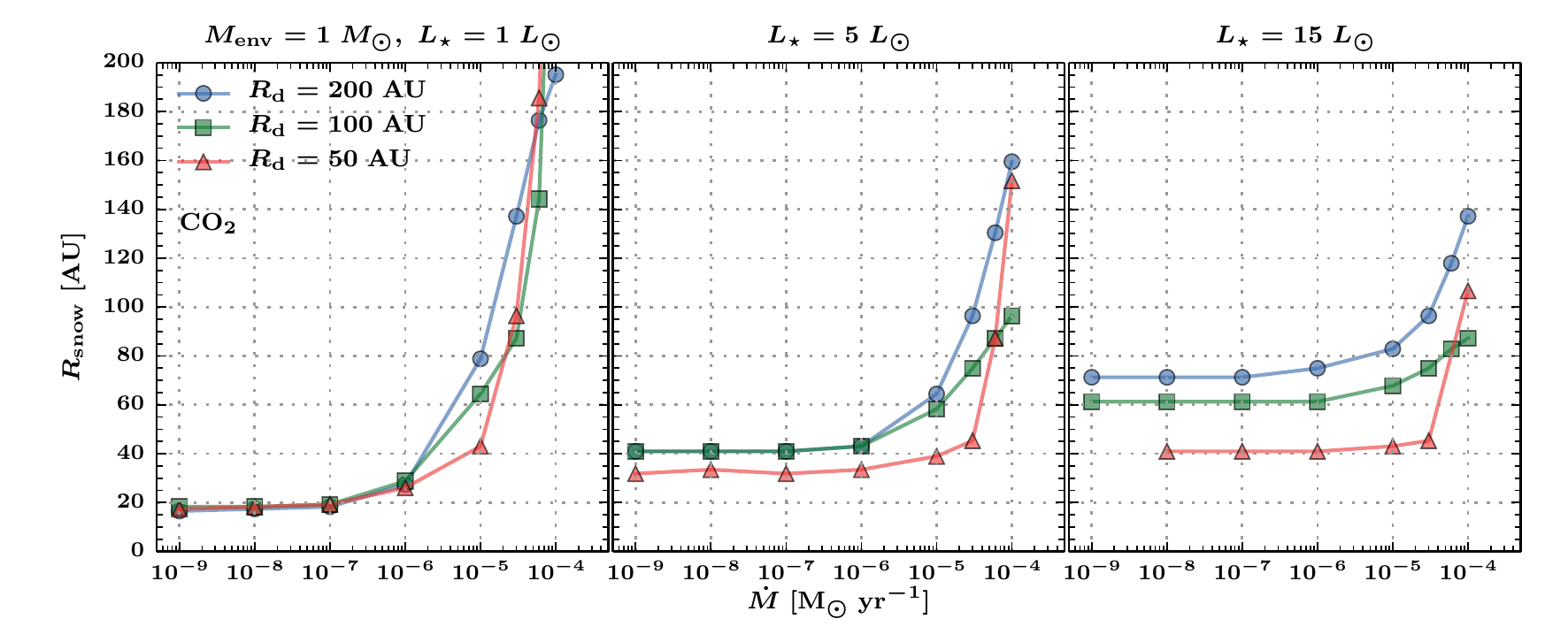}
\caption{Midplane CO$_2$ snowline as a function of stellar 
luminosity, accretion rate, and disk radius.  The parameters are 
similar to that of Fig.~\ref{fig:giratio3}. }
\label{fig:giratio3co2}
\end{figure*}

\begin{figure*}[tpb]
\centering
\includegraphics{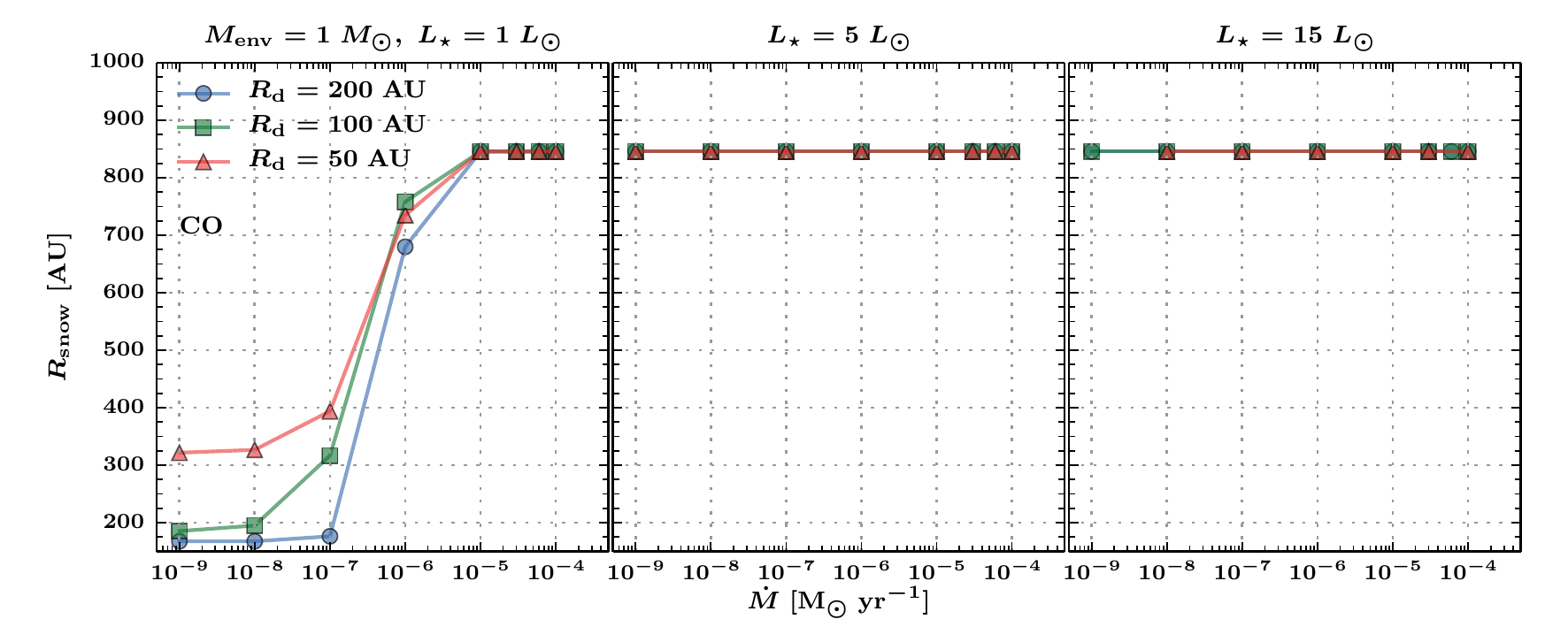}
\caption{Midplane CO snowline as a function of stellar luminosity, 
accretion rate, and disk radius.  The parameters are 
similar to that of Fig.~\ref{fig:giratio3}.}
\label{fig:giratio3co}
\end{figure*}

The H$_2$O, CO$_2$ and CO gas pressure dependent gas fraction abundance 
are shown in Fig.~\ref{fig:gasice}.  Figures~\ref{fig:giratio3co2} and 
\ref{fig:giratio3co} show the midplane CO$_2$ and CO snowlines as 
a function of luminosity and $R_{\rm d}$ similar to that of 
Fig.~\ref{fig:giratio3} for H$_2$O. 

\end{document}